\documentstyle[preprint,prb,aps,epsf]{revtex}

\begin{document}
\preprint{SU-ITP \#95/32, cond-mat/9512009 }
\tighten

\title{A new class of collective excitations of the Hubbard model I:
\mbox{$\eta$ excitation} of the negative--$U$ model}

\author{Eugene Demler\thanks{e-mail address: \tt
eugene@quantum.stanford.edu} and Shou-Cheng Zhang}
\address{Department of Physics, Stanford University, Stanford, CA~~94305}

\author{Nejat Bulut and Douglas J. Scalapino}
\address{Department of Physics, UC at Santa Barbara,
Santa Barbara, CA~~93106}

\date{\today}

\maketitle

\begin{abstract}

In this series of papers we present a detailed study of the
particle--particle collective excitations of the Hubbard model, and their
contribution to the density and spin excitation spectrum.
In the first paper,
we shall investigate the singlet particle--particle pair
with momentum $(\pi,\pi)$,
the $\eta$ particle, of the negative--$U$ Hubbard model.
We review three
previously obtained theorems about the $\eta$ particle and
develop a self-consistent linear response theory which
takes into account its contribution to the density excitation
spectrum in the superconducting state.
We show that this self--consistent theory
agrees with the exact theorems as well as the
results of numerical Monte Carlo simulations.

\end{abstract}

\pacs{74.72.Bk, 61.12.Bt, 61.12.Ex}

\section{INTRODUCTION}
\label{sec:intro}

The Hubbard model provides a basic framework within which the nature
of strongly--correlated electron systems have been studied.
Recently, it was found that the Hubbard model has $SO(4)$
symmetry\cite{Yang-Zhang}, for both positive and
negative on-site Coulomb energy
$U$. That is, besides the usual $SU(2)$ spin-rotation symmetry,
it is also invariant under an $SU(2)$ ``pseudo--spin''
rotation group which contains the charge $U(1)$ symmetry
as a subgroup. Based on this new symmetry, one of us
\cite{Zhang1,Zhang2} derived a number of theorems and argued that this
symmetry implies the existence of a collective excitation,
the $\eta$--mode.
This mode is characterized by a charge number of two, a spin quantum
number zero, total momentum $(\pi,\pi)$
and a sharp energy $U-2\mu$.
Here $\mu$ the chemical potential.
As the momentum shifts away from $(\pi,\pi)$, the mode
disperses and eventually merges into the continuum.
Now, since this excitation has a charge quantum number of two,
one can not couple to it in the normal state with the
usual electric or magnetic--field
probes\footnote{In principle, a two--particle tunneling process could couple
to this excitation but this would also require a superconductor.}
used to study the density and spin excitations of a many body system.
However, if the system were to become superconducting, spontaneously
breaking the $U(1)$ charge symmetry, then one should observe this
$\eta$ mode in the charge density response.
Thus one has the unusual situation, that above $T_c$
there is a well defined excitation in the many--body system
which is invisible to the usual probes and it is only below $T_c$ that
we see this excitation.

Here we explore this phenomena for the negative--$U$ Hubbard model.
We begin with a review of the exact theorems regarding the $\eta$--mode.
We then use a self--consistent linear response theory to approximately
calculate correlations of the $\eta$ and the charge density $\rho$ in the
superconducting state.
We show that this approach gives results in agreement with the
exact theorems and hence provides a meaningful approximation.
Finally, we carry out Monte Carlo simulations for the two--dimensional
negative--$U$ Hubbard model and compare these with the approximate solution.
In the Monte Carlo case, we can calculate the $\eta$--$\eta$
correlation function above the Kosterlitz--Thouless temperature
and clearly see the sharp $\eta$--mode in the normal state in this
correlation function.  We can then calculate the $\rho$--$\rho$
charge density response and see that the $\eta$--mode becomes visible
in the charge density response when the temperature is lowered and the
pairing correlations extend over the lattice.
We discuss these results and their possible relevance to other
problems in the conclusions.

The formalism of self-consistent linear response theory takes into
account the coupling of the particle-hole excitation and
particle-particle excitation in the superconducting state. Such a
formalism was first developed in the context of gauge invariance
problem in superconductivity\cite{Anderson1,Rickayzen,Schrieffer}.
It was later used by Bardasis and
Schrieffer\cite{Bardasis,vanderMarel} to discuss the possibility of
excitonic states
inside the BCS energy gap. While the formalism we use here is similar to
these previous works, the physical origin of the $\eta$ mode is very
different from the excitonic states inside the BCS energy gap.
The excitonic states owe their existence to the BCS energy gap, they
exist near a total momentum near zero, and disappear completely
above $T_c$. The $\eta$ mode, on the other hand, owes its existence
to the collapse of the particle particle continuum near $(\pi,\pi)$
and exists even in the normal state, although they decouple from the
density spectrum above $T_c$. After the prediction of the $\eta$ mode
in the Hubbard model\cite{Zhang1}, Kostyrko and Micnas\cite{Kostyrko2}
have applied the conserving approximation in Nambu formalism to study
the collective modes of the extended Hubbard model at zero
temperature, and numerically confirmed the existance of the $\eta$
mode in their formalism. In our present work we shall make the
detailed comparison of $SO(4)$ Ward identities and the correlation
functions derived using the self-consistent linear response
theory, invstigate the effect of finite temperature on the $\eta$ mode,
and demonstrate explicitely the scaling of its intensity with the
superconducting order parameter.

Our present investigation is motivated in part by the recent neutron
scattering experiments on the high $T_c$ superconductors, in which a
collective resonance was found below the superconducting transition
temperature\cite{Grenoble1,Grenoble,Mook1,Keimer}.
This mode was interpreted by two of us\cite{pi} as a triplet
particle particle collective mode, hereafter called the $\pi$ mode,
of the positive $U$ Hubbard model.
Our motivation to study the details of the $\eta$ mode of the
negative $U$ Hubbard model is to use it
as a theoretical laboratory to verify a mechanism in which a particle
particle collective mode can couple to the particle hole spectrum below
the superconducting transition temperature, and to check the methodology
used in the calculations. The $\eta$ and $\pi$ modes both have
charge two and exist near momentum $(\pi,\pi)$, but they have different
spin quantum numbers.
The $\eta$ mode has spin zero and energy of the order of $|U|$, while
the $\pi$ mode has spin one and energy of the order of $J$, the spin
exchange energy.
The similarities and
the differences of these two collective modes will be addressed in detail
in our next paper.

\section{REVIEW OF THREE EXACT THEOREMS}
\label{sec:theorem}

In this section we shall review three main theorems of the Hubbard
model derived by one of us in references\cite{Zhang1,Zhang2}. These
results follow from Yang's work\cite{Yang} on the $\eta$ pairing, and the
$SO(4)$ symmetry of the Hubbard model\cite{Yang-Zhang}. However, it
is important to point out a physical difference between
the following discussion and the idea of $\eta$ pairing. While the
idea of $\eta$ pairing\cite{Machida,Yang} refers to a ground state
property of the model,
the following discussion assumes a conventional, zero total momentum
pairing in the ground state, and the $\eta$ pair is considered as
an excitation of the system.

Let us start with the Hubbard Hamiltonian defined by
\begin{eqnarray} {\cal H} =
  -t \sum_{<ij>}{{c^{\dagger}_{i\sigma} c_{j\sigma}}}
  +U \sum_{i}{{ n_{i\uparrow} n_{i\downarrow}}}
  -\mu \sum_{i}{{c^{\dagger}_{i\sigma} c_{i\sigma}}}
\label{model}
\end{eqnarray}
Within this model, one can define the following operators
\begin{eqnarray}
\eta^\dagger & = & \sum_{p} c_{p+Q\uparrow}^\dagger
c_{-p\downarrow}^\dagger \nonumber \\
\eta & = & (\eta^\dagger)^\dagger \nonumber \\
\eta_0 & = & \frac{1}{2} (N_e - N)
\end{eqnarray}
where $Q=(\pi,\pi)$, $N_e$ is the total number of electrons and
$N$ is the total number of lattice sites. It is easy to see that
they form an $SU(2)$ algebra, called the pseudospin algebra:
\begin{equation}
[\eta_0, \eta^\dagger] = \eta^\dagger\ , \ [\eta_0, \eta] = -\eta\ , \
[\eta^\dagger, \eta] = 2 \eta_0
\end{equation}
These $\eta$ operators have a remarkable commutation property with the
Hubbard Hamiltonian
\begin{equation}
[H, \eta^\dagger] = (U-2\mu) \eta^\dagger\ , \ [H, \eta_0] = 0\ , \
[H, \eta^2] = 0
\label{w0}
\end{equation}
Since the Hubbard Hamiltonian commutes with the Casimir operators
$\eta_0$ and $\eta^2 \equiv \frac{1}{2} (\eta^\dagger \eta + \eta \eta^\dagger)
+ \eta_0 \eta_0$ of the pseudospin algebra, the Hubbard Hamiltonian
posses an $SO(4)=SU(2)_{spin}\times SU(2)_{pseudospin}$ symmetry group.
These commutation rules and the symmetry property can be used to derive
the following three important theorems, initially obtained in
\cite{Zhang1,Zhang2}.

First let us consider the pure particle--particle correlation function
involving only the $\eta$ operators:
\begin{eqnarray}
P(q,\omega) & = & - i \int dt e^{i\omega t}< T \eta_q (t) \eta_q^\dagger
(0) > \nonumber\\
&  = & \sum_{n} \left\{
  \frac{ \left| \langle n | \eta^{\dagger}_q | 0 \rangle \right|^2 }
  { \omega - E_{n0} + i 0}
-  \frac{ \left| \langle n | \eta_q | 0 \rangle \right|^2 }
  { \omega + E_{n0} - i 0} \right\}
\label{particle-particle}
\end{eqnarray}
where $\eta_q^\dagger= \sum_{p} c_{p+q\uparrow}^\dagger
c_{-p\downarrow}^\dagger$ and the summation in the second line is  over
all the excited states $|n\rangle$.

 For the case of $q=Q$, the operator
$\eta_q$ reduces to the $\eta$ operator defined above. Throughout this
paper we shall assume without loss of generality that system is less
than half-filled, i.e. $N_e < N $. From the
commutation relation with the Hubbard Hamiltonian, it is simple to
see that $\eta^{\dagger}$ operator creates a single excited state when acting
on the ground state,\footnote { In the consequent formulas
we will be using the filling factor $n = N_e/N$ rather than $N_e$ and
will be omitting all the factors of $N$ for brevity. So, that
summation over all momenta $p$ in our notations gives $1$ and not
$N$. This may be viewed as analog of normilization by a unit
volume in the continious case. When needed the explicit dependence on
$N$ may be easily recovered.}
\begin{equation}
| \eta \rangle = ( 1 - n )^{ - \frac{1}{2}} ~\eta^{\dagger} | 0
\rangle
\end{equation}
where a normalization factor is introduced such that
\begin{eqnarray}
\langle \eta | \eta \rangle & = & ( 1 - n )^{-1} \langle 0 | \eta~
\eta^{\dagger} | 0 \rangle =  ( 1 - n )^{-1} \langle 0 | \left[
\eta , \eta^{\dagger} \right] | 0 \rangle \nonumber\\ & = & ( 1
- n )^{-1} \sum_{p} ( 1-n_{p\downarrow}-n_{p+Q\uparrow} ) = 1
\label{pp}
\end{eqnarray}

It is also easy to show that $ \eta $ annihilates the
vacuum: $ \eta | 0 \rangle = 0 $ (we have
already used this fact in (\ref{pp}) to replace the product of $ \eta$
and $ \eta^{\dagger} $ by their commutator).

Therefore only one term survives in (\ref{particle-particle})
with $\langle n | = \langle \eta | $. And for $P(Q,\omega)$ we have
\begin{eqnarray}
P(Q,\omega) = \frac{1}{ ( 1 - n )}
\frac{\left|  \langle 0 | \eta~ \eta^{\dagger} | 0 \rangle \right|^2}
{ \omega - \omega_0 + i 0}
= \frac{  ( 1 - n ) }{ \omega - \omega_0 + i 0}
\label{Pres}
\end{eqnarray}
where $\omega_0 = U - 2 \mu $ is the energy of the $\eta$-particle.
Equation (\ref{Pres}) is the content of our first theorem.

Next let us consider a mixed correlation function involving the
$\eta$ operator and the density operator
$ \rho_{q} =  \sum_{k\sigma} c^{\dagger}_{ k+q \sigma } c_{
  k \sigma } $,
\begin{eqnarray}
M(q,\omega) & = & - i \int dt e^{i\omega t}< T \rho_{-q} (t) \eta_q^\dagger
(0) > \nonumber\\
& = & \sum_{n} \left\{
  \frac{  \langle 0 | \rho_{-q} | n \rangle
    \langle n | \eta^{\dagger}_q | 0 \rangle}
  { \omega - E_{n0} + i 0}
- \frac{  \langle 0 | \eta^{\dagger}_q| n \rangle
    \langle n | \rho_{-q}| 0 \rangle}
  { \omega + E_{n0} - i 0} \right\}
\label{mixed}
\end{eqnarray}
For $q=Q$ we have contribution from the $|\eta\rangle$-state only,
and $M(Q,\omega)$ becomes
\begin{equation}
M(Q,\omega) = \frac{ \langle 0 | \rho_{-q}  \eta_q^\dagger
 | 0 \rangle }{ ( 1 - n ) } \frac{ 1 } { \omega - \omega_0 + i 0} = -
2 \frac{ < \Delta > }{ U  ( 1 - n ) }
\frac{ 1 }{ \omega - \omega_0 + i 0}
\label{Mres}
\end{equation}
where $\Delta = - U \sum_k  c_{k\uparrow}^{\dagger}
c_{-k\downarrow}^{\dagger} $
is the superconducting order parameter.
Equation (\ref{Mres}) is our second theorem.

Finally, let us consider the pure density--density correlation function
defined by
\begin{eqnarray}
D(q,\omega) & = & - i \int dt e^{i\omega t} < T \rho_{-q} (t) \rho_q
(0) >  \nonumber\\
& = &\sum_{n} \left| \langle 0 | \rho_{-q} | n \rangle \right|^2
\left\{ \frac{ 1 }{ \omega  - E_n + i 0} - \frac{ 1 }{ \omega + E_n -
    i 0} \right\}
\label{density}
\end{eqnarray}
For $D(Q,\omega)$ we can pull out the part singular at $\omega_0$
\begin{eqnarray}
D(Q,\omega) & = & \frac{ \left| \langle 0 | \rho_{-Q} | \eta \rangle
\right|^2 } { \omega - \omega_0 + i 0} + part~regular~at~\omega_0
\nonumber\\
& = & ( 1 - n )^{-1} \frac{ \left| \langle 0 | \left[
   \rho_{-Q} , \eta^{\dagger} \right] | 0 \rangle
\right|^2 } { \omega - \omega_0 + i 0} + part~regular~at~\omega_0 \nonumber\\
& = & 4 \frac{ { < \Delta > }^2 }{ U^2 ( 1 - n ) }~ \frac{ 1 }{\omega
  - \omega_0 + i 0} + part~regular~at~\omega_0
\label{Dres}
\end{eqnarray}
Equation (\ref{Dres}) is our third theorem.

{}From expressions (\ref{Pres}), (\ref{Mres}) and (\ref{Dres}) we can
see that all the correlation functions have poles at $\omega_0$ and
their residues are expressed as simple combinations of parameters of
the system. We want to emphasize again that these three results are
exact and do not require any approximations in their derivations,
and they are valid for both the positive and the negative $U$
Hubbard model.

Among all three of these theorems, the third one is the most interesting,
since the density correlation function can be measured directly
in experiments. This theorem predicts that
a new collective excitation, the $\eta$ particle
can be found in the density spectrum,
whose intensity is directly proportional to the
square of the superconducting
order parameter, in this case, an $s$--wave order parameter.
Physically this is because the density correlation function measures
a particle hole like excitation, and therefore, the $\eta$ particle,
which has charge two rather than zero, can only make a finite
contribution if the
quasi--particle excitations are
a coherent admixture of particles
and holes. The mixing amplitude is exactly proportional to the square of
the superconducting order parameter.
For a momentum $Q=(\pi,\pi)$ the energy of the
$\eta$--mode is  $\omega_0=U-2\mu$.
However, unlike the first two theorems, the third theorem
only makes a partial prediction about a correlation function, namely
the part that is singular at $\omega_0$. For this reason, it is
highly desirable to find an approximate scheme which yields complete
information about the full momentum and frequency range of the
density correlation function, and reduces to the exact theorem for the
special momentum $Q$ and frequency $\omega_0$. Such an investigation
is carried out in the subsequent sections.

\section{SELF-CONSISTENT LINEAR RESPONSE THEORY}
\label{sec:FSLRT}
\subsection{Formalism}

In calculating the density response in the superconducting state,
it is known that one must take into account the collective motion
of the condensate in a self--consistent manner as well as the usual
quasi--particle--hole excitations.
An externally applied density disturbance
$\phi_q$ at wave length $q$ generates two kinds of responses in the system,
single--particle like excitations across the energy gap, and
a collective motion of the condensate expressed as a
paring amplitude $\eta_q$, and a density
field $\rho_q$, at a finite wave length $q$.
The usual RPA formula in the superconducting state
takes into account the self-consistent density field $\rho_q$,
but does not treat the self-consistent pairing field $\eta_q$. As we
shall show, such an approximation is not fully self-consistent and
violates the $SO(4)$ symmetry of the Hubbard model. On the other
hand, a fully self-consistent treatment of both the density and the
pairing field correctly describes the contribution of the $\eta$
particle in the superconducting state, and reduces to the exact
theorems reviewed in the previous section.

In the following, we review the basic formalism.
In the Appendix we compare
our results to the formulas one gets from the equations of motion method
and from the diagrammatical approach.
In both cases we find exact agreement.

Our starting point is the negative--$U$ Hubbard Hamiltonian perturbed by an
infinitesimally small external field $\phi_q$ which is coupled to the
$-q$ component of the density.
\begin{eqnarray} {\cal H}  =
  -t \sum_{<ij>}{{c^{\dagger}_{i\sigma} c_{j\sigma}}}
  +U \sum_{i}{{ n_{i\uparrow} n_{i\downarrow}}}
  -\mu \sum_{i}{{c^{\dagger}_{i\sigma} c_{i\sigma}}}
  +\frac{1}{2}U \rho_{-q}\phi_{q}
\label{lrmodel}
\end{eqnarray}
We will assume that our system is in a superconducting ground
state characterized by the nonvanishing averages of the zero momentum
operators
\begin{eqnarray}
\langle c^{\dagger}_{k\sigma}
c_{k\sigma} \rangle & = & v_k^2 \nonumber \\
\langle c_{-k\downarrow} c_{k\uparrow}
\rangle =
\langle c^{\dagger}_{k\uparrow} c^{\dagger}_{-k\downarrow}
\rangle & = & u_k v_k.
\label{0averages}
\end{eqnarray}
Here the parameters $u_k$ and $v_k$ are given by
\begin{eqnarray}
u_k^2=\frac{1}{2} (1+\frac{\epsilon_k}{E_k}) \\
v_k^2=\frac{1}{2} (1-\frac{\epsilon_k}{E_k})
\end{eqnarray}
with $\epsilon_k = - 2 t ( \cos k_x + \cos k_y ) - \mu$ , $E_k =
\sqrt{\epsilon_k^2 + \Delta^2}$
and the BCS self-consistency condition is implied
\begin{equation}
\Delta = - U \sum_p u_p v_p
\end{equation}
After we turn on the perturbation, the operators in
(\ref{0averages}) are no longer the only operators with nonvanishing
expectation values. The expectation values of the following operators
carrying momentum $q$ are induced as well.
\begin{eqnarray}
\langle \eta^{\dagger}_q(t)\rangle & = & \sum_{k}\langle~
c^{\dagger}_{k+q\uparrow} (t)
c^{\dagger}_{-k\downarrow} (t)~\rangle \nonumber\\
\langle \eta_q(t)\rangle  & = & \sum_{k}\langle~ c_{-k-q\downarrow} (t)
c_{k\uparrow} (t)~\rangle \nonumber\\
\langle \rho_{q\uparrow}(t) \rangle & = & \sum_{k}\langle~
c^{\dagger}_{k+q\uparrow} (t)
c_{k\uparrow} (t)~\rangle \nonumber\\
\langle \rho_{q\downarrow}(t)\rangle  & = & \sum_{k}\langle~
c^{\dagger}_{-k\downarrow} (t)
c_{-k-q\downarrow} (t)~\rangle
\label{fields}
\end{eqnarray}
The operators in (\ref{fields}) describe the collective motion of the
superconducting
condensate. Within the self-consistent linear response theory, the
local operators respond to the local field, which is the sum of the
external disturbance $\phi_q$ and the induced self-consistent field
given in equation (\ref{fields}).
It is very important that in (\ref{fields}) we take into account not
only the particle-hole channel as is usually done in RPA but also
the particle-particle and hole-hole channels.

The Hamiltonian (\ref{lrmodel}) can be linearized with respect to the
usual BCS pairing fields and the induced fields (\ref{fields})
giving \footnote{ To prevent any possible confusion we want to clarify
that in this Section
$\eta_q $, $\rho_{q \uparrow \downarrow}$ and $\eta_q^{\ast}= \langle
 \eta^{\dagger}_q  \rangle$ are the { \it expectation values }
of the corresponding operators and in commutation relations must be
considered as c-numbers.} :
\begin{eqnarray} {\cal H}  & =  &
  \sum_{p\sigma}{\epsilon_p c^{\dagger}_{p\sigma} c_{p\sigma}}
  +\Delta c^{\dagger}_{p\uparrow} c^{\dagger}_{-p\downarrow}
  +\Delta c_{-p\downarrow} c_{p\uparrow}\nonumber\\
   & +  & U \eta^{\ast}_q \sum_{k'} c_{-k'\downarrow}c_{k'+q \uparrow}
  + U \eta_q \sum_{k'}
c^{\dagger}_{k'-q\uparrow}c^{\dagger}_{-k'\downarrow}\nonumber\\
   & +  & U ( \rho_{q\uparrow}+\frac{\phi_q}{2}) \sum_{k'}
  c^{\dagger}_{k'-q\downarrow} c_{k'\downarrow}
  +U ( \rho_{q\downarrow}+\frac{\phi_q}{2}) \sum_{k'}
  c^{\dagger}_{k'-q\uparrow} c_{k'\uparrow}
\label{lham}
\end{eqnarray}
The first line in the last equation will be considered as an
unperturbed Hamiltonian ${\cal H}_{0}$. Since the fields in (\ref{fields})
are proportional to $\phi_q$ one can use the Kubo formula
and treat the last four terms in (\ref{lham}) as a perturbation ${\cal
  H}_1$.
Thus  the response $\langle \hat{f}(t)~\rangle$ is given by
\begin{equation}
\langle \hat{f}(t)~\rangle= -i \int_{-\infty}^{t}
dt' {\langle ~\left[~ \hat{f}(t), {\cal H}_{1}(t')~\right]~
\rangle}_{{\cal H}_0},
\label{self-cons}
\end{equation}
and $f$ can be any of the operators $\eta^{\dagger}_q$, $\eta_q$,
$\rho_{q\uparrow}$ or $\rho_{q\downarrow}$.
When combined, these
equations form a self-consistent set of equations and the
expectation values in equation (\ref{fields}) can be solved
for purely in terms
of the external field $\phi_q$.
As an example
of how one can proceed from (\ref{self-cons}) we
will show explicitly the calculations for $\eta^{\ast}_q$.
\begin{eqnarray}
\eta^{\ast}_{q}(t) = & -  & i \sum_k \int_{-\infty}^{t} dt' {\langle~
\left[~ c^{\dagger}_{k+q\uparrow}(t) c ^{\dagger}_{-k\downarrow}(t),
{\cal H}_{1}(t')~\right]~\rangle}_{{\cal H}_0}= \nonumber\\
 & - &  i U \int_{-\infty}^{t} \eta^{\ast}_{q}(t') dt' \sum_{kk'}
{\langle~
\left[~
c^{\dagger}_{k+q\uparrow}(t) c ^{\dagger}_{-k\downarrow}(t),
c_{-k'\downarrow}(t')
c_{k'+q\uparrow}(t')~\right]~\rangle}_{{\cal H}_0} \nonumber\\
 & -  & i U \int_{-\infty}^{t} \eta_{q}(t') dt' \sum_{kk'}
{\langle~
\left[~
c^{\dagger}_{k+q\uparrow}(t) c ^{\dagger}_{-k\downarrow}(t),
c^{\dagger}_{k'-q\uparrow}(t')
c ^{\dagger}_{-k'\downarrow}(t')~\right]~\rangle}_{{\cal H}_0} \nonumber\\
 & -  & i U \int_{-\infty}^{t} (\rho_{q\uparrow}(t')+\frac{1}{2}\phi_q(t'))
dt' \sum_{kk'}
{\langle~ \left[~
c^{\dagger}_{k+q\uparrow}(t) c ^{\dagger}_{-k\downarrow}(t),
c^{\dagger}_{k'-q\downarrow}(t')
c_{k'\downarrow}(t')~\right]~\rangle}_{{\cal H}_0} \nonumber\\
 & -  & i U \int_{-\infty}^{t} (\rho_{q\downarrow}(t')+\frac{1}{2}\phi_q(t'))
dt' \sum_{kk'}
{\langle~ \left[~
c^{\dagger}_{k+q\uparrow}(t) c ^{\dagger}_{-k\downarrow}(t),
c^{\dagger}_{k'-q\uparrow}(t')
c_{k'\uparrow}(t')~\right]~\rangle}_{{\cal H}_0}
\label{b+time}
\end{eqnarray}
When we perform a Fourier transform of the last equation it is helpful to
make use of the usual trick of replacing the commutator with a
time-ordered Green's
function and shifting the poles to the lower-half plane at the
end. This will give us the following expression for $\eta_{q\omega}
\equiv \int dt
\eta_{q}(t) e^{ i\omega t}$ ( Fourier components of the other functions
are defined analogously )
\begin{eqnarray}
\eta^{\ast}_{q\omega}= & i  &  U \sum_p \int \frac{d \nu}{2 \pi} G_p(\nu)
G_{p+q}
(-\nu-\omega) \eta^{\ast}_{q\omega}\nonumber\\
-  & i  &  U \sum_p \int \frac{d \nu}{2 \pi} F^*_p(\nu) F^*_{p+q}
(\nu+\omega) \eta_{q\omega}\nonumber\\
- & i &  U \sum_p \int \frac{d \nu}{2 \pi} F^*_p(\nu) G_{p+q}
(\nu-\omega) ( \rho_{q\omega} + \phi_{q\omega})
\label{b+freq}
\end{eqnarray}
where we introduced
$\rho_{q\omega}=\rho_{q\omega\downarrow}+\rho_{q\omega\uparrow}$, and
$ G_p(\omega) \equiv \int e^{i \omega t} ( - i ) \langle
T c_{p\sigma}(t) c^{\dagger}_{p\sigma}(0)  \rangle dt $ and $ F_p(\omega)
\equiv \int e^{ i \omega t} ( - i ) \langle
T c_{p\uparrow}(t) c_{-p\downarrow}(0) \rangle dt $ are the usual BCS
Greens functions. We remind the reader
that after performing the frequency integration in (\ref{b+freq})
one should shift all the poles of the final expression to the lower
half-plane. Similar expressions can be easily obtained for $
\eta_{q\omega}$ and $\rho_{q\omega}$.
\begin{eqnarray}
\eta_{q\omega}=-  & i &  U \sum_p \int \frac{d \nu}{2 \pi} F_p(\nu) F_{p+q}
(+\nu+\omega) \eta^{\ast}_{q\omega}\nonumber\\
+ & i  & U  \sum_p \int \frac{d \nu}{2 \pi} G_p(\nu) G_{p+q}
(\omega-\nu) \eta_{q\omega}\nonumber\\
- & i  & U \sum_p \int \frac{d \nu}{2 \pi} F_p(\nu) G_{p+q}
(\nu+\omega) ( \rho_{q\omega} + \phi_{q\omega})
\label{b-freq}
\end{eqnarray}

\begin{eqnarray}
\rho_{q\omega}=- & i  & U \sum_p \int \frac{d \nu}{2 \pi} F_{p+q}(\nu) G_p
(\nu-\omega) \eta^{\ast}_{q\omega}\nonumber\\
-  & i  & U \sum_p \int \frac{d \nu}{2 \pi} F^*_{p+q}(\nu) G_p
(\nu+\omega) \eta_{q\omega}\nonumber\\
+ & i  & U \sum_p \int \frac{d \nu}{2 \pi} \left\{ F_p(\nu) F_{p+q}
(\nu+\omega) - G_p (\nu) G_{p+q}(\nu+\omega) \right\}
( \rho_{q\omega} + \phi_{q\omega})
\label{rofreq}
\end{eqnarray}

In the diagrammatic language these
equations correspond to the vertex corrections for scattering in
the particle-particle, particle-hole and hole-hole channels and are
presented in Appendix B.
Generalization of formulas (\ref{b+freq}) - (\ref{rofreq}) to the case of
finite temperatures can be accomplished by substituting \mbox{$i \int
  \frac{d\nu}{2\pi} \rightarrow -T\sum_{\nu_n}$}. Skipping laborious
but straightforward calculations we present the
final expressions leaving their derivation aside. We introduce the
matrix equation
\begin{eqnarray}
\left(\begin{array}{c}
\eta^{\ast}_{q\omega}+\eta_{q\omega} \\
\eta^{\ast}_{q\omega}-\eta_{q\omega} \\
\rho^{\dagger}_{q\omega}
\end{array} \right)
=\left| \begin{array}{ccc}
t_{++} & t_{+-} & m_+ \\ t_{+-} & t_{--} & m_- \\ m_+ & m_- & -U \chi_{bcs}
\end{array} \right|
\left(\begin{array}{c}
\eta^{\ast}_{q\omega}+\eta_{q\omega} \\
\eta^{\ast}_{q\omega}-\eta_{q\omega} \\
\rho_{q\omega}+\phi_{q\omega}
\end{array} \right)
\label{matrix}
\end{eqnarray}
where the matrix elements are given by the following expressions,
\begin{eqnarray}
t_{++} & =  & U \sum_{p}
 \left[1-f(E_{p+q})-f(E_p)\right]\frac{\Omega_{pq}
( 1-v_{p+q}^2-v_{p}^2)}
{\omega^2-\nu_{pq}^2 +i 0} \nonumber\\
 & +  & U \left[f(E_{p+q})-f(E_p)\right]
\frac{(E_{p+q}-E_p)(u_p v_{p+q}+v_p u_{p+q})^2}
{\omega^2-\theta_{pq}^2 +i 0}
\label{t++}\\
  t_{+-}  & =  & -U \sum_{p}
 \left[1-f(E_{p+q})-f(E_p)\right]\frac{\omega
( 1-v_{p+q}^2-v_{p}^2)}
{\omega^2-\nu_{pq}^2 +i 0} \nonumber\\
 & -  & U \left[f(E_{p+q})-f(E_p)\right]
\frac{\omega( v_{p+q}^2 - v_p^2)}
{\omega^2-\theta_{pq}^2 +i 0}
\\
t_{--} & =  & U \sum_{p}
 \left[1-f(E_{p+q})-f(E_p)\right]\frac{\nu_{pq}
( u_{p+q} u_p+v_{p+q} v_p)^2}
{\omega^2-\nu_{pq}^2 +i 0} \nonumber\\
 & + & U \left[f(E_{p+q})-f(E_p)\right]
\frac{\theta_{pq} (u_p v_{p+q}-v_p u_{p+q})^2}
{\omega^2-\theta_{pq}^2 +i 0}
\\
m_{+}  & =  & U \sum_{p}
 \left[1-f(E_{p+q})-f(E_p)\right]\frac{\Omega_{pq}
( u_p v_p+u_{p+q} v_{p+q})}
{\omega^2-\nu_{pq}^2 +i 0} \nonumber\\
 & -  &U \left[f(E_{p+q})-f(E_p)\right]
\frac{\Omega_{pq}(u_{p+q} v_{p+q}-u_p v_p)}
{\omega^2-\theta_{pq}^2 +i 0}
\\
m_{-}  &=  & -U \sum_{p}
 \left[1-f(E_{p+q})-f(E_p)\right]\frac{\omega
( u_p v_p+u_{p+q} v_{p+q})}
{\omega^2-\nu_{pq}^2 +i 0} \nonumber\\
 & + & U \left[f(E_{p+q})-f(E_p)\right]
\frac{\omega (u_{p+q} v_{p+q}-u_p v_p)}
{\omega^2-\theta_{pq}^2 +i 0}
\\
\chi_{bcs} & =  & - \sum_{p}
 \left[1-f(E_{p+q})-f(E_p)\right]\frac{\nu_{pq}
( u_{p+q} v_p+v_{p+q} u_p)^2}
{\omega^2-\nu_{pq}^2 +i 0} \nonumber\\
 & +  & \left[f(E_{p+q})-f(E_p)\right]
\frac{\theta_{pq} (u_p u_{p+q}-v_p v_{p+q})^2}
{\omega^2-\theta_{pq}^2 +i 0}
\label{chi}
\end{eqnarray}
In the equations above $\omega_{pq}= \epsilon_{p+q} - \epsilon_p$ ,
$\Omega_{pq}= \epsilon_{p+q}+\epsilon_p$ , $\nu_{pq}= E_{p+q}+E_p$ ,
$\theta_{pq}= E_{p+q}-E_p$ and $f(E)$ is the Fermi distribution
function.

Solution of the density part of the matrix equation (\ref{matrix}) can
be written in a very suggestive form:
\begin{equation}
\rho_{q\omega} = \frac{-U \chi_{bcs} +
\frac{m_{+}^{2}+m_{-}^{2}-m_{-}^{2}t_{++}+2 m_{+} m_{-} t_{+-}-m_{+}^2 t_{--}}
{1-t_{++}-t_{+-}^2-t_{--}+t_{++}t_{--}}}
{1+U \chi_{bcs}-
\frac{m_{+}^{2}+m_{-}^{2}-m_{-}^{2}t_{++}+2 m_{+} m_{-} t_{+-}-m_{+}^2 t_{--}}
{1-t_{++}-t_{+-}^2-t_{--}+t_{++}t_{--}}
} ~\phi_{q\omega}
\label{resp}
\end{equation}
It is easy to see that the last terms in both the numerator and the
denominator are proportional to the BCS pairing amplitude, therefore,
this expression reduces to the usual RPA formula in the normal state.
In the superconducting state however, these correction terms make
important contributions, and it would be {\it incorrect} to neglect them.
These terms take into account
the self-consistent pairing fields in the superconducting state, and
in the diagrammatic language, they arise from multiple scattering in
the particle--particle and the hole--hole channels.
Equation (\ref{resp}) agrees with the expression for the
density-density correlation function derived by Kostyrko and
Micnas\cite{Kostyrko2}.
Formulas (\ref{matrix}) - (\ref{resp}) are the
core of the Self-consistent Linear Response Theory.
In Part B we will show
analytically that for the case of $q=Q$ and $\omega=\omega_0$,
these formulas reduce to the exact theorem given
in the previous section, with the correct location of the poles
and the residues.
In Part C we compare these results with numerical calculations.

\subsection{Comparisons with exact theorems}
\label{subsec:comparison}
Here we show how our formulas (\ref{matrix}) for
$T=0$ reduce to the exact results (\ref{Mres}) and (\ref{Dres}) for
$Q=(\pi,\pi)$. This point has a property that $\Omega_{pQ}= -2 \mu$
for all momenta $p$. Therefore $\Omega_{pQ}$ may be taken outside the $p$
summation in the equations of (\ref{matrix}) and a number of
interesting identities arise.

We start by multiplying the first equation in (\ref{matrix}) by
$\omega$, the second by
$\Omega_{pq}$ and adding these two equations. After a few simple
manipulations the resulting expression can be written  as
\footnote { We remind the reader that since the present analysis is for
 $T=0$, all the fermi functions in (\ref{t++}) - (\ref{chi}) vanish. }
\begin{equation}
\omega ( \eta^{\ast}_{Q\omega}+ \eta_{Q\omega} )= \left [ - \Omega_{pQ}-
U ( 1 - n ) \right ] ( \eta^{\ast}_{Q\omega}- \eta_{Q\omega} )
\label{id1}
\end{equation}

We can prove the following
relationship
 $
\nu_{pq} (v_p u_{p+q}+v_{p+q} u_p)^2  =
\left[\frac{\nu_{pq}^2-\omega^2}{2 \Delta} +
\frac{\omega^2-\Omega_{pq}^2}{2 \Delta} \right] (u_p v_p + u_{p+q}
v_{p+q})
$
and use it to write the third equation of (\ref{matrix}) in the form

\begin{eqnarray}
-2 \Delta \phi_{q\omega}=  & 2 & \Delta U \sum_{p}
\frac{\Omega_{pq} (u_p v_p+u_{p+q} v_{p+q} )}
{\omega^2-\nu_{pq}^2 +i 0} (\eta^{\ast}_{q\omega}+\eta_{q\omega}) \nonumber\\
- & 2  & \Delta U \sum_{p}
\frac{\omega (u_p v_p+u_{p+q} v_{p+q})}
{\omega^2-\nu_{pq}^2 +i 0} (\eta^{\ast}_{q\omega}-\eta_{q\omega}) \nonumber\\
+ & U  &\sum_{p}
\frac{\omega^2-\Omega_{pq}^2}
{\omega^2-\nu_{pq}^2 +i 0}(u_p v_p + u_{p+q} v_{p+q})
(\rho_{q\omega}+\phi_{q\omega})
\label{D}
\end{eqnarray}

We add to (\ref{D}) the first equation of (\ref{matrix}) multiplied by
$\Omega_{pQ}$ and the second multiplied by $\omega$ to
obtain\footnote{ Equations (\ref{id1}) and (\ref{id2}) arise
 directly if one writes the equations of motion for $\sum_p
 ( \eta^{\ast}_{pq} - \eta_{pq} )$ and $\sum_p ( \eta^{\ast}_{pq}
 + \eta_{pq} )$}

\begin{equation}
\omega ( \eta^{\ast}_{Q\omega}- \eta_{Q\omega} ) =
\left [ - \Omega_{pQ}- U ( 1 - n ) \right ] ( \eta^{\ast}_{Q\omega} +
  \eta_{Q\omega} ) + 2 \Delta \phi_Q
\label{id2}
\end{equation}
Equations (\ref{id1}) and (\ref{id2}) can be solved for
$\eta^{\ast}_{Q\omega}$ and $ \eta_{Q\omega}$ in terms of $\phi_q$
\begin{equation}
\eta^{\ast}_{Q\omega}= \frac{ \Delta} { \omega + \omega_0}
\phi_{Q\omega}
\hspace{4cm}
\eta_{Q\omega}= - \frac{ \Delta} { \omega - \omega_0}
\phi_{Q\omega}
\label{omega0}
\end{equation}
where we introduced $ \omega_0 = U ( 1 - n ) + \Omega_{pQ} = U ( 1 - n
) - 2 \mu $.\footnote{ This expression may seem to give a different
  value for $\omega_0$ than (\ref{w0}). This comes from the fact that
  in the Linear Response we do not take into account the Hatree-Fock
  corrections to the self-energy or in other words in the diagrams of
  Appendix C we do not consider corrections to the single particle
  Green's functions. In fact, the only effect of these corrections is
  to renormalize a chemical potential by an average effective
  field on each particle due to the interaction with the particles of
  opposite spin $\mu \longrightarrow \mu - \frac{U n}{2}$. After such
  substitution this last expression for
  $\omega_0$ is exactly the same as in (\ref{w0}). This procedure of
  Hartree-Fock corrections reducing to the renormalization of $\mu$ is
  explicit in the equations of motion.}
Now, the expressions (\ref{omega0}) by themselves give us a mixed
correlation function (\ref{mixed}) or they can be inserted into
any of the equations of (\ref{matrix}) giving the density-density
correlation function
\begin{eqnarray}
M ( Q, \omega ) & =  & \frac { \eta_{q\omega}}{\frac{U}{2} \phi_{q\omega}}
= - \frac{ 2 \Delta} { U } \frac {1}{ \omega - \omega_0 + i 0}
\\
D ( Q, \omega )  & = & \frac { \rho_{q\omega}}{\frac{U}{2} \phi_{q\omega}}
= - 2 \frac{ \chi_{bcs}(Q,\omega) + 2 \frac{\omega^2 - 2 \mu
    \omega_0} {\omega^2 - \omega_0^2}  \Delta I_2(Q,\omega)} { 1 + U
  \chi_{bcs} (Q,\omega)}
\label{Dexpr}
\end{eqnarray}
where $I_2(Q,\omega)= \sum_p \frac{ u_p v_p
  + u_{p+Q} v_{p+Q}}{\omega^2-\nu_{pQ}^2} $. From (\ref{Dexpr}) we
immediately see that $ D(Q,\omega)$ has a pole at  $\omega=\omega_0$
corresponding to the $\eta$-pair and using
\begin{equation}
1+ U
\chi_{bcs}(Q,\omega) = - \frac{ U }{2 \Delta } ( \omega^2 -
\Omega_{pQ}^2) I_2(Q, \omega)
\end{equation}
we can write (\ref{Dexpr}) as
\begin{equation}
D(Q,\omega) = 4 \frac{\Delta^2}{U^2 (1 - n )}~ \frac{1}{\omega-\omega_0
  +i 0 } + part~ regular~ at~ \omega_0
\end{equation}
Thus the Self-consistent Linear Response
expressions agree with the exact theorems, giving
the correct position of the resonance
as well as the same values for the residues of all
correlation functions.
Here we have explicitly verified theorems 2 and
3 from last section. In order to verify theorem 1, one needs to add
an external pairing disturbance, in addition to the density disturbance.
Repeating the manipulations given above, it can be seen that theorem
1 is verified as well.

These results show that while the Self-consistent Linear Response
Theory is still an approximate scheme, it is conserving in the sense that
it respects the $SO(4)$ Wards identities expressed through theorems 1 to
3. On the other hand, a simple RPA expression in which one
neglects the last terms in the numerator and the denominator of
equation (\ref{resp}) violates the $SO(4)$ symmetry and is a poor
approximate scheme.

\subsection{Numerical results}
\label{sec:num}
Here we give numerical results for the spectral functions
calculated from equation (\ref{resp}), with the full
momentum, energy and temperature dependence. The calculations
have been carried out for different $U$s' and $n$'s and the values of
parameters are specified
in the captions for each figure.
The imaginary infinitesimal in the denominators was
taken to be $\Gamma = 1.0 \times 10^{-2}$
and the integrations were performed on
a $1024 \times 1024$ lattice.

In figure (\ref{ LR + RPA + BCS }) we plot the imaginary part of the
density response function, namely the spectral function, at zero
temperature and $q=Q$. In curve
(a) we show the spectral function calculated from the Self-consistent
Linear Response theory, and show that the location of the peak agrees
exactly with the value predicted by the exact theorem, indicated by a
delta function arrow. The width of the peak is not intrinsic, but
due to the finiteness of $\Gamma$ used to smooth the calculations.
In curve (b) we show the results from the usual RPA formula, and we
see that it completely misses the location of the collective mode.
For a larger value of $U$, this discrepancy becomes even stronger. In curve
(c) we show the result of the simple BCS
quasiparticle approximation to the
density response function. It has an onset at the minimal value of
$E_p + E_{p+Q}=2|\mu|$, but no sharp peak structure of the collective mode.

In figures (\ref{dispfigU1}) and (\ref{dispfigU4}) we show the
momentum dependence of the peak, by plotting
the spectral function for various values of
the center of mass momentum.
It is interesting to note the difference in behaviour of the peak
for different values of the interaction strength $U$. For the case of
smaller $U$'s shown in Figure (\ref{dispfigU1}),
we see that the collective mode is most sharply defined at $Q=(\pi,\pi)$,
and that
it broadens and disperses downwards in energy as one moves away from
$Q$. For larger $U$'s, however, the energy of the mode increases
as we go away from $(\pi,\pi)$ with only a small change in the width
or intensity of the peak. Within a simple T-matrix approximation of
the $\eta$-mode, i.e. without the mixing into the density excitations,
one can show that the $\eta$-mode always disperses downwards as one
moves away from the $(\pi,\pi)$ point, and it merges { \it
  tangentially }into the particle-particle continuum. The upward
dispersion of the $\eta$-mode for large U can be attributed to the
strong mixing of the $\eta$-mode with density excitations. We present
a more detailed discussion of the dispersion of the mode and
comparison between T-matrix approxiamtion and the Self-consistent
Linear Response method in Section
\ref{sec:mapping}.

 It is important to remind the reader that the special properties
of the momentum $Q$ demonstarted in figures above have nothing to do
with the nesting property of the
fermi surface, since we are studying a doped system where $2k_f$ is
markably different from $Q$ at this filling. Instead, the special
role of $Q$ is that at this momentum, the $\eta_q$ operator becomes
an eigenoperator of the Hamiltonian. Physically speaking, this is the
momentum where the particle--particle continuum vanishes.

In figure (\ref{ Tfig }) we plot the temperature dependence of the
spectral function for momentum $Q$. We see that the peak intensity
decreases with temperature and the peak vanishes exactly at $T_c$. This
behavior is in exact agreement with the third theorem that the
peak intensity is proportional to the BCS order parameter. We also see
that at low temperatures, the weight of the collective mode is
transferred from the higher energy continuum.

\vspace{ -2.5 cm}
\def\ifundefined#1{\expandafter\ifx\csname#1\endcsname\relax}
\begin{figure}[hbt]
  \ifundefined{epsffile}\relax\else
\centerline{\epsfysize 9cm \epsffile[18 140 592 698]{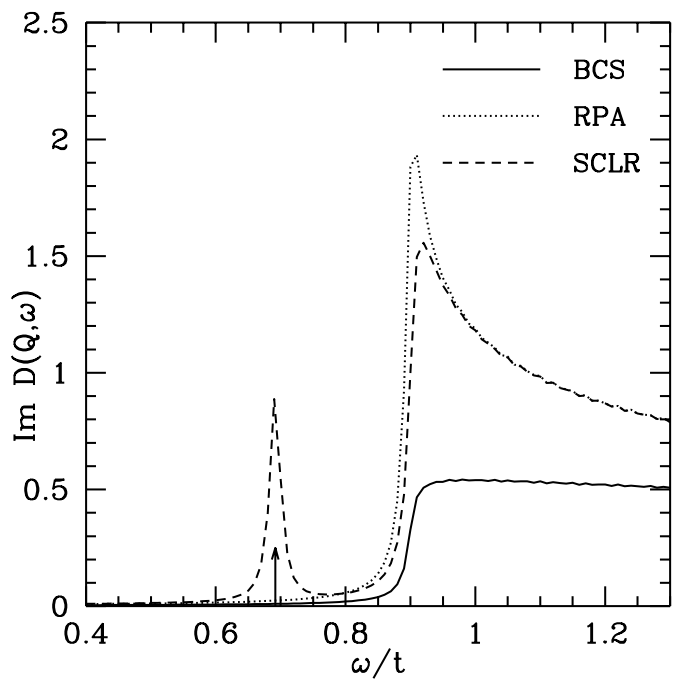}}\fi
\caption{  $Im D (Q, \omega)$ versus $\omega$ calculated using
  BCS, Random Phase Approximation and Self-consistent Linear Response
  Theory formulas. The arrow indicates $\omega_0$ - a
  position of the $\eta$-pair as predicted by the exact theorems. This
  plot is for $U = -t$, $n=0.79$ and $T = 0$. }
\label{ LR + RPA + BCS }
\end{figure}

\begin{figure}[hbt]
    \ifundefined{epsffile}\relax\else
\centerline{\epsfysize 9cm \epsffile[18 140 592 698]{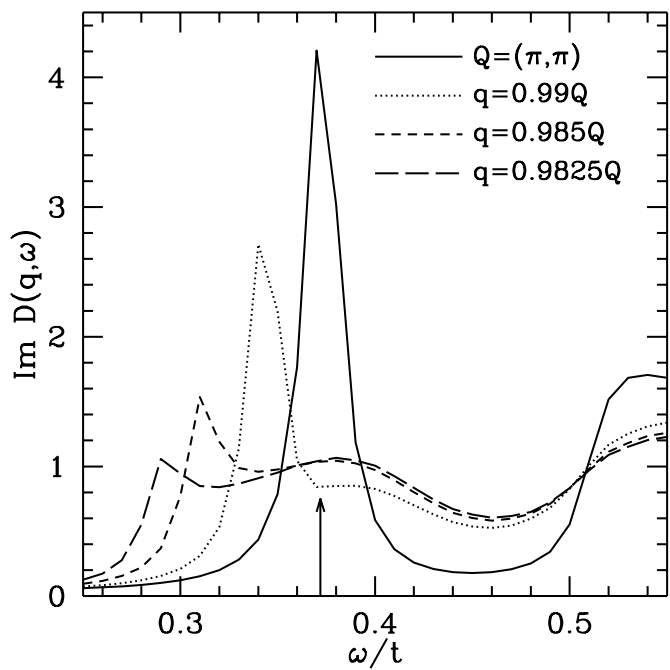}}\fi
\caption{  $Im D(q, \omega)$ versus $\omega$ for $U = -t$, $n=0.87$, $T
  = 0$ and different momenta $q$.
The arrow marks $\omega_0 $. This figure shows that for small $U$'s
the collective $\eta$-mode exists only in
a very close vicinity of $(\pi,\pi)$. One can also see a small dispersion
of the mode - as we go away from $(\pi,\pi)$ the energy decreases. }
\label{dispfigU1}
\end{figure}

\begin{figure}[hbt]
    \ifundefined{epsffile}\relax\else
\centerline{\epsfysize 9cm \epsffile[18 140 592 698]{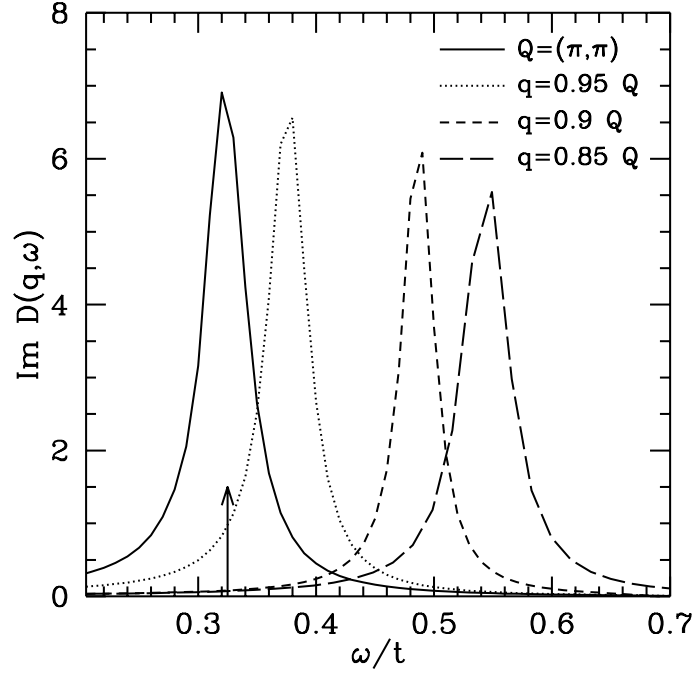}}\fi
\caption{ $Im D(q, \omega)$ versus $\omega$ for $U = -4 t$, $n=0.87$,
$T = 0$ and  different momenta $q$.
The arrow marks $\omega_0 $. Notice a big positive dispersion and
absense of strong damping of the mode away from
$(\pi,\pi)$.  }
\label{dispfigU4}
\end{figure}

\begin{figure}[hbt]
    \ifundefined{epsffile}\relax\else
\centerline{\epsfysize 9cm \epsffile[18 140 592 698]{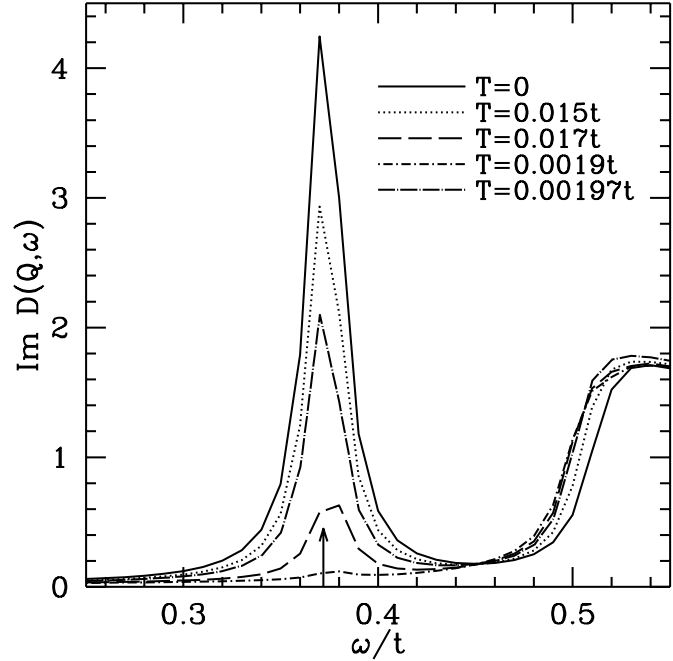}}\fi
\caption{ $Im D (Q, \omega)$ versus $\omega$ for $ U = - t $, $ n =
  0.87 $ and different
  temperatures:  $ T = 0$ corresponds to $\Delta = 0.041 t $;
 $ T = 0.015 t$ to $\Delta = 0.031 t $;   $ T = 0.017 t$ to
$\Delta = 0.021 t $;
$ T = 0.019 t$ to $ \Delta = 0.012 t $ and $ T = 0.0197 t$ to $ \Delta
= 0.002 t $.  The arrow marks $\omega_0 $. It is interesting that as we
raise the temperature, the intensity of the peak decreases but its
position does not change. }
\label{ Tfig }
\end{figure}

\section{MONTE CARLO CALCULATIONS}
\label{sec:monte-carlo}

It is useful to compare the results of the exact theorems and
the self--consistent calculations with the numerical data from
Quantum Monte Carlo simulations.
Using the finite--temperature determinantal Monte Carlo technique\cite{QMC}
we have measured the Matsubara time--ordered Green's functions
\begin{equation}
D(q,\tau)=<T_{\tau} \rho_{-q}(\tau) \rho_q(0) >
\label{Dqt}
\end{equation}
and
\begin{equation}
P(q,\tau)=<T_{\tau} \eta_{q}(\tau) \eta^{\dagger}_q(0) >.
\label{Pqt}
\end{equation}
The spectral weights ${\rm Im}\,D(q,\omega)$ and
${\rm Im}\,P(q,\omega)$ can be obtained by maximum entropy analytic
continuation \cite{maxent1,maxent2}
of $D(q,\tau)$ and $P(q,\tau)$ to the real frequency axis.

In this section, we will present results for $U=-4t$ and $-1t$
on an $8\times 8$ lattice.
We will consider electron fillings $\langle n\rangle =0.87$,
0.7 and 0.5.
For a filling of $0.87$ and $U=-4t$,
the Kosterlitz--Thouless superconducting transition temperature
$T_c^{\rm KT}$ is $\simeq 0.1t$ \cite{negU1,negU2}.
In this parameter regime, we have carried out calculations for
temperatures varying from $T=0.5t$ down to $0.08t$.

The solid line in Fig. (\ref{ImP1}) shows ${\rm Im}\,P(q,\omega)$
at center--of--mass momentum $q=(\pi,\pi)$ for $U=-4t$,
$\langle n\rangle =0.87$ and $T=0.33t$.
Here the chemical potential is
$\mu=-2.15t$.
We see that the peak is centered at $\omega_0\simeq 0.37t$,
which is close to $U-2\mu$.
The finite width of the peak is due to the resolution of the maximum
entropy method.
The dotted and the dashed curves in Fig. (\ref{ImP1})
show ${\rm Im}\,P(q,\omega)$ for $q=(\pi,3\pi/4)$ and
$(3\pi/4,3\pi/4)$.
We observe that the $\eta$ resonance broadens and shifts to higher
frequencies as $q$ moves away from $(\pi,\pi)$.
The excitation is expected to disperse as
$|q-(\pi,\pi)|^2$,
however, the momentum resolution on the $8\times 8$ lattice
is $\pi/4$, and hence these calculations can't provide information on the
dispersion of the $\eta$-resonance in the immediate vicinity of
$(\pi,\pi)$.

Fig. (\ref{ImD1}) shows the spectral weight of the density
response function, ${\rm Im}\,D(q,\omega)$,
at $q=(\pi,\pi)$ for various values of $T$.
As the temperature is lowered a sharp peak developes
at low frequencies.
We note that the position of the peak in
${\rm Im}\,D(q=(\pi,\pi),\omega)$ coincides with the position of the
peak in ${\rm Im}\,P(q=(\pi,\pi),\omega)$.
In the grand canonical ensemble, the chemical potential
depends on the temperature, and
this is reflected in a slight shift of the peak position.

We can obtain further information on the density fluctations
by studying $D(q,\omega=0)$.
Fig. ({\ref{Dq}) shows $D(q,0)$ versus $q$ for $q$ around the Brillouin
zone at various temperatures.
According to the Kramers--Kronig relation
\begin{equation}
D(q,0) = \int_0^{\infty} {d\omega'\over \pi}
{ {\rm Im}\,D(q,\omega') \over \omega'},
\label{KK}
\end{equation}
$D(q,0)$ provides us with information on the integral
of the spectral weight divided by $\omega$.
In this figure we see that $D(q=(\pi,\pi),0)$
increases rapidly as $T$ is lowered.
We believe that this rapid growth of $D(q=(\pi,\pi),0)$ is due to the
coupling of the density excitations to the $\eta$ particle--particle
channel as $T\rightarrow T_c^{\rm KT}$.
However, for $q\neq (\pi,\pi)$ such a rapid growth is not observed.
This would then mean that the $\eta$ excitation is relatively local in
momentum space.

In Figure (\ref{ImD1}), we have seen that the peak in
${\rm Im}\,D$  starts growing for temperatures considerably
higher than $T_c^{\rm KT}$.
This is possibly due to the fact that there are significant
superconducting fluctuations above $T_c^{\rm KT}$.
In order to obtain information on the strength of these superconducting
fluctuations, we consider the current--current correlation function
\begin{equation}
\Lambda_{xx}(q,i\omega_m) = {1\over N} \int_0^\beta d\tau
e^{i\omega_m \tau}
< j_x(q,\tau) j_x(-q,0) >
\label{Lambda}
\end{equation}
with
\begin{equation}
j_x(q) = it\sum_{\ell}
e^{-iq\cdot\ell}
(c_{\ell+x \sigma}^{\dagger} c_{\ell \sigma} -
c_{l \sigma}^{\dagger} c_{\ell+x \sigma}).
\label{jx}
\end{equation}
The superfluid density $D_s$ is given by \cite{SWZ}
\begin{equation}
{D_s\over \pi e^2} =
-\langle k_x\rangle
-\Lambda_{xx}(q_x=0,q_y\rightarrow 0,i\omega_m=0)
\label{Dsup}
\end{equation}
where $\langle k_x\rangle$ is the average kinetic energy
in the $x$ direction.
In Fig. (\ref{Ds}) we have plotted
$-\langle k_x\rangle
-\Lambda_{xx}(q_x=0,q_y,i\omega_m=0)$
versus $q_y$ at various temperatures.
These results, which are identical to those given in
Fig. (8) of Ref. \cite{SWZ}, show that there are significant
superconducting fluctuations on the $8\times 8$ system
for $T$ higher than $T^{\rm KT}_c\simeq 0.1t$.
Basically the coherence length of the superconducting fluctuations
is extending over the finite $8\times 8$ lattice.
We also note that the growth of $D_s$ and the peak
in ${\rm Im}\,D$ appear to be strongly correlated,
as one would have expected from the exact theorems of Section II.

For $U=-4t$ the statistical error in the Quantum Monte Carlo
data is large, and as a result of this the spectral weights get
smeared by the maximum entropy analytic continuation.
For instance, in Fig. (\ref{ImP1}) we have seen that the
$\eta$ resonance, which is a $\delta$ function,
was broadened considerably, and its frequency was shifted.
However, for $U=-1t$ we can obtain Monte Carlo data with
better statistics and the resulting spectral weights
have improved resolution.
Unfortunately, for $U=-1t$, the superconducting
transition takes place at temperatures much lower than it is
feasible for a numerical study.
Hence, we will only study the $\eta$ pair field susceptibility
at the elevated temperature of $T=0.25t$.

Here we consider the filling dependence of the $\eta$ resonance.
Fig. (\ref{ImP3}) shows ${\rm Im}\,P(q=(\pi,\pi),\omega)$ versus
$\omega/t$ for fillings of 0.87, 0.7 and 0.5.
The corresponding chemical potentials are $-2.16t$,
$-2.67t$ and $-3.25t$.
We see from this figure that the resonance frequencies are
very close to the expected values of $U-2\mu$.
In addition, we find that the total spectral weights in the peaks are
0.14, 0.31 and 0.49, respectively.
These values agree well with the exact theorem of Eq. (8),
which gives $1-\langle n\rangle$ for the weight.

In this section, we have seen that a sharp resonance exists in the
$\eta$--pairing channel.
As the temperature is lowered and the superconducting correlations
develop across the $8\times 8$ lattice, the density fluctuations
can couple to this channel.
The resulting density excitation spectrum has a peak at the resonance
frequency in addition to a continuum of excitations.
These results are in agreement with the predictions
of the exact theorems discussed in Section II and the results
of the self--consistent calculations presented in Section III.

\begin{figure}[hbt]
    \ifundefined{epsffile}\relax\else
\centerline{\epsfysize 10cm \epsffile[18 184 592 598]{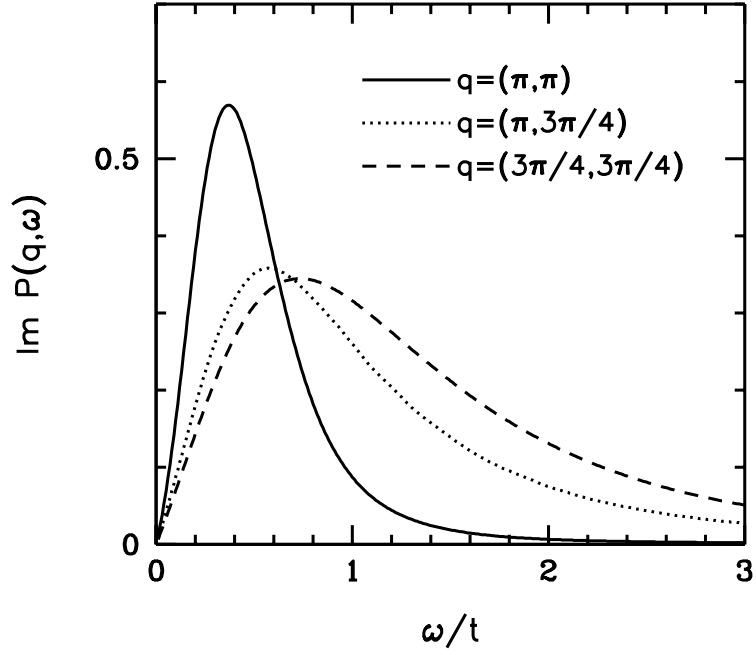}}\fi
\caption{${\rm Im}\,P(q,\omega)$ versus $\omega/t$
at various values of $q$
for $U=-4t$, $\langle n\rangle =0.87$ and $T=0.33t$.}
\label{ImP1}
\end{figure}

\begin{figure}[hbt]
    \ifundefined{epsffile}\relax\else
\centerline{\epsfysize 10cm \epsffile[18 184 592 598]{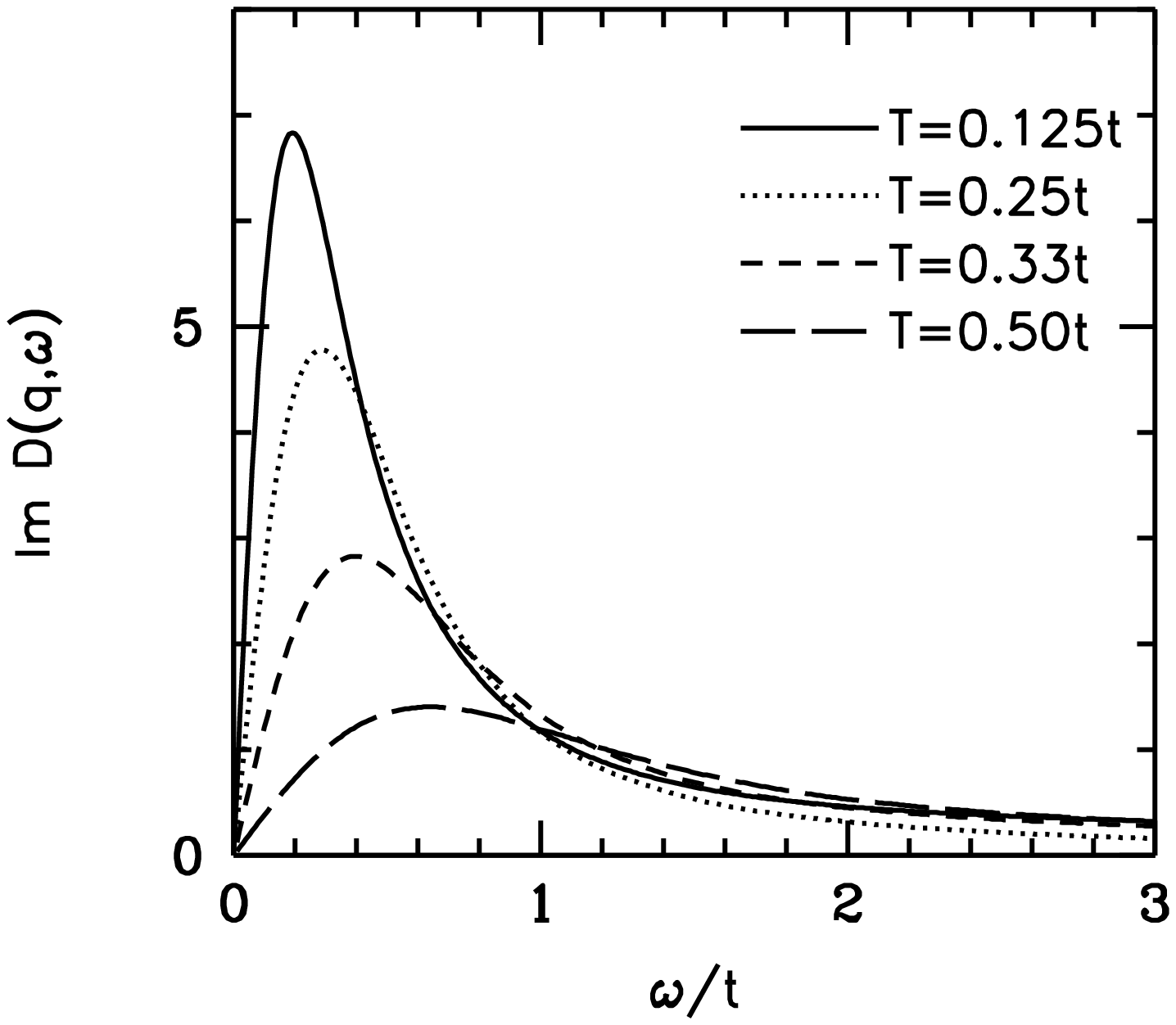}}\fi
\caption{${\rm Im}\,D(q=(\pi,\pi),\omega)$ versus $\omega/t$ for
$U=-4t$, $\langle n\rangle =0.87$ and various temperatures.}
\label{ImD1}
\end{figure}

\begin{figure}[hbt]
    \ifundefined{epsffile}\relax\else
\centerline{\epsfysize 10cm \epsffile[18 184 592 598]{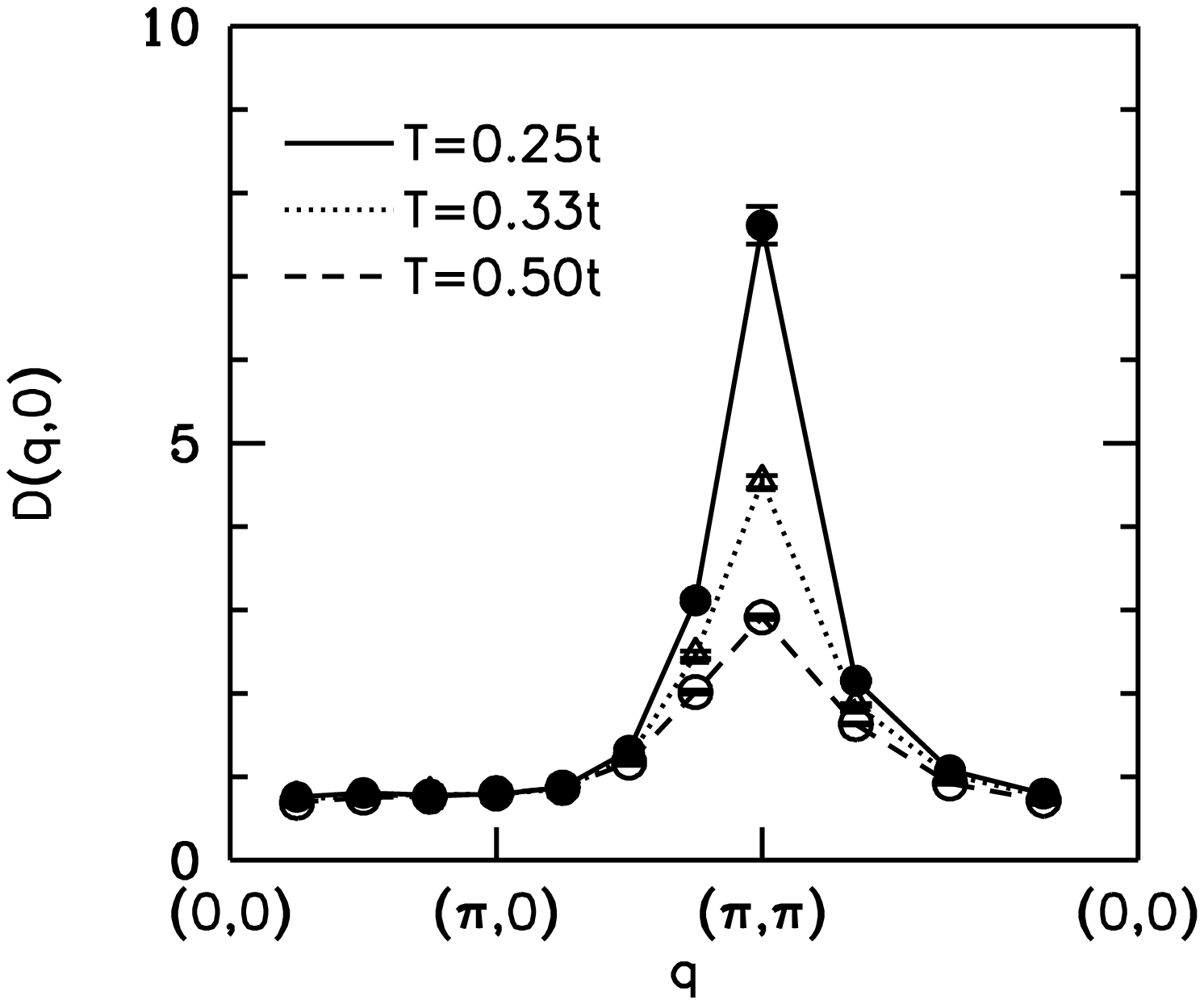}}\fi
\caption{$D(q,\omega=0)$ versus $q$ for
$U=-4t$, $\langle n\rangle =0.87$ and
various temperatures.}
\label{Dq}
\end{figure}

\begin{figure}[hbt]
    \ifundefined{epsffile}\relax\else
\centerline{\epsfysize 10cm \epsffile[18 174 592 598]{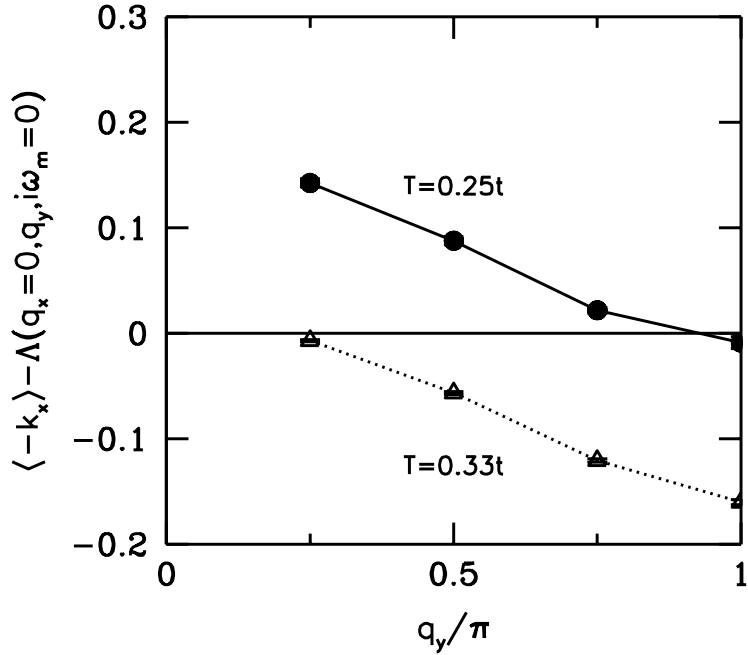}}\fi
\caption{$\langle -k_x\rangle - \Lambda_{xx}(q_x=0,q_y,i\omega_m=0)$
versus $q_y/\pi$ for
$U=-4t$, $\langle n\rangle =0.87$ and various temperatures.
The $q_y\rightarrow 0$ limit of this quantity gives the superfluid
density, $D_s/\pi e^2$.}
\label{Ds}
\end{figure}

\begin{figure}[hbt]
    \ifundefined{epsffile}\relax\else
\centerline{\epsfysize 10cm \epsffile[18 184 592 598]{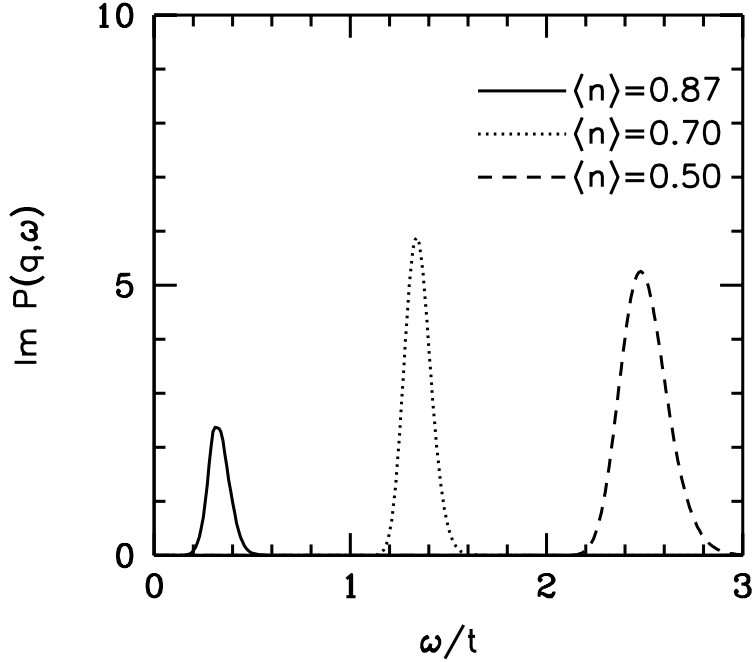}}
\caption{${\rm Im}\,P(q=(\pi,\pi),\omega)$ versus $\omega/t$
for $U=-1t$, $T=0.25t$
and various values of $\langle n\rangle$}
\label{ImP3}
\end{figure}

\section{MAPPING OF THE $\eta$-MODE TO ANTIFERROMAGNETIC SPIN WAVE}
\label{sec:mapping}

\subsection{Particle-hole transformation}
\label{sec:transf}

There is an elegant way to derive the dispersion of the $\eta$-mode
for the negative $U$ Hubbard model in the large $|U|$ limit
by mapping it onto a Heisenberg model in an external magnetic
field. Mele first used this formalism to discuss the dispersion of the
$\eta$ mode for zero temperature\cite{Mele}.
Below we generalize this approach to include the effects of finite
temperature and obtain the temperature dependence of the $\eta$ mode
dispersion.

Our starting point is a particle-hole
transformation\cite{Shiba,Hirsch,negU2} on the bipartite lattice:

\begin{eqnarray}
c_{ i \uparrow} &\rightarrow& c_{ i \uparrow} \nonumber\\
c_{  i \downarrow} &\rightarrow& \left\{ \begin{array} {l}
+ c_{ i \downarrow}^{\dagger},~ i \in A \\
- c_{ i \downarrow}^{\dagger},~ i \in B \end{array} \right.
\label{transf}
\end{eqnarray}
where A and B stands for the two sublattices.
In the future we will write the transformation of the spin-down
particles in the form
\mbox{ $ c_{  i \downarrow} \rightarrow (-)^i
c_{i\downarrow}^{\dagger} $}.

For the two-fermion operators (\ref{transf}) gives

\begin{eqnarray}
c_{i \uparrow}^{\dagger} c_{i \uparrow} & \rightarrow &
c_{i \uparrow}^{\dagger} c_{i \uparrow} \nonumber\\
c_{i \downarrow}^{\dagger} c_{i \downarrow} & \rightarrow &
c_{i \downarrow} c_{i \downarrow}^{\dagger} =
1 - c_{i \downarrow}^{\dagger} c_{i \downarrow}
\label{quad}
\end{eqnarray}
and so the charge $ Q = n_{\uparrow} + n_{\downarrow} $
becomes the spin  $ S_z = \frac{1}{2} ( n_{\uparrow} - n_{\downarrow}
) $ and vice versa.

\begin{eqnarray}
Q \rightarrow 2 S_z +1 \hspace{2cm} 2 S_z \rightarrow Q-1
\end{eqnarray}
If we now write our hamiltonian in the form

\begin{eqnarray}
{\cal H } = -t \sum_{<ij> \sigma} c_{i \sigma}^{\dagger} c_{j \sigma}
- | U | \sum_i (n_{i \uparrow}-\frac{1}{2}) (n_{i
  \downarrow}-\frac{1}{2}) - \tilde{\mu} \sum_{ i \sigma} c_{i
  \sigma}^{\dagger} c_{i \sigma}
\end{eqnarray}
which is different from the earlier notation by only a chemical
potential renormalization $ \tilde{ \mu } = \mu + \frac{| U | n}{2} $,
 we obtain using (\ref{quad})
\begin{eqnarray}
{\cal H } \rightarrow {\cal H } = -t \sum_{<ij> \sigma} c_{i
  \sigma}^{\dagger} c_{j \sigma} + | U | \sum_i (n_{i
  \uparrow}-\frac{1}{2}) (n_{i \downarrow}-\frac{1}{2}) - 2 \tilde{\mu}
\sum_{ i \sigma} S_{iz}
\label{negU}
\end{eqnarray}

Hamiltonian (\ref{negU}) describes a positive U Hubbard model at
half-filling \footnote{ We refer the reader to a well known theorem
  that $\mu = \frac{U}{2}$ at half-filling } in the presence of
a magnetic field $2\tilde{\mu}$ \footnote{ Here and
  everywhere else we take Bohr's magneton to be one, $\mu_B=1$.},
which in the large $U$ limit is
known to be equivalent to the Heisenberg model
\begin{eqnarray}
{\cal H} = J \sum_{<ij>} {\bf S}_i \cdot {\bf S}_j - h \sum_{i} S_{i z}
\label{heisenberg}
\end{eqnarray}
with $J=\frac{4t^2}{U}$ and $h = 2 \tilde{\mu}$. Therefore by
considering the Heisenberg model, we will be
able to get a deeper insight into the modes of the negative $U$
Hubbard model.

Our next step will be to establish the
correspondence between excitations of the two systems.
To do so, we notice that the particle-hole transformation turns
$SU(2)_{pseudospin}$ algebra into a usual $SU(2)_{spin}$ algebra

\begin{eqnarray}
\eta^{\dagger} = \sum_i (-)^i c_{i \uparrow}^{\dagger} c_{ i
  \downarrow}^{\dagger} \hspace{1cm} &\rightarrow&  \hspace{1cm} S_{+}=
\sum_i c_{i\uparrow}^{\dagger} c_{i \downarrow} \nonumber\\
\eta = \sum_i (-)^i c_{i \downarrow} c_{ i
  \uparrow} \hspace{1cm} &\rightarrow&  \hspace{1cm} S_{-}=
\sum_i c_{i\downarrow}^{\dagger} c_{i \uparrow} \nonumber\\
\eta_0 = \frac{1}{2} ( N_e - N )
\hspace{1cm} &\rightarrow&  \hspace{1cm}
S_z = \frac{1}{2} \sum_i ( c_{i \uparrow}^{\dagger} c_{i \uparrow} -
c_{i \downarrow}^{\dagger} c_{i \downarrow} )
\label{SU(2)}
\end{eqnarray}

And it transforms Cooper pairs and charge density wave into
antiferromagnetic waves in $x-y$ and $z$ directions
respectively.

\begin{eqnarray}
\Delta^{\dagger} = \sum_i c_{i \uparrow}^{\dagger} c_{i
  \downarrow}^{\dagger} \hspace{1cm} &\rightarrow& \hspace{1cm} S_{+}(Q)
= \sum_i (-)^i  c_{i \uparrow}^{\dagger} c_{i \downarrow} \nonumber\\
\Delta = \sum_i c_{i \downarrow} c_{i
  \uparrow} \hspace{1cm} &\rightarrow& \hspace{1cm} S_{-}(Q)
= \sum_i (-)^i  c_{i \downarrow}^{\dagger} c_{i \uparrow} \nonumber\\
\rho(Q) = \sum_i \left( c_{i \uparrow}^{\dagger} c_{i \uparrow}  +
c_{i \downarrow}^{\dagger} c_{i \downarrow} \right)
 \hspace{1cm} &\rightarrow& \hspace{1cm}
2 S_z(Q) = \sum_i (-)^i \left( c_{i \uparrow}^{\dagger} c_{i \uparrow}  -
c_{i \downarrow}^{\dagger} c_{i \downarrow} \right)
\label{AF}
\end{eqnarray}

Hamiltonian (\ref{heisenberg}), as well as any other
rotationally invariant system subject
to an external magnetic field, has a Larmor precession mode given by
\begin{eqnarray}
\left[ {\cal H}, S_{\pm} \right] = \pm h S_{\pm}
\end{eqnarray}
So, from (\ref{SU(2)}) we see that $\eta$-excitation is what becomes
of the Larmor motion under the particle-hole transformation.

Coupling of the $\eta$ excitation to the density fluctuations also has
its analogy in the Larmor motion.
We assumed $s$-wave superconductivity in the negative $U$
case, then after the particle-hole transformation takes us to the
Heisenberg model we must have an antiferromagnetic order in $x-y$
plane or non-vanishing $ <S_{\pm}(Q)>$. The BCS order
parameter being real implies that antiferromagnetism will be in
$x$-direction. This means that $ \left[ S_{\pm}, S_z(Q) \right] =
S_{\pm}(Q) $  is a c-number and so the
Larmor motion is conjugate to the $S_z(Q)$ oscillations.
Thus when we
excite Larmor oscillations we will have a response in $S_z(Q)$.
This is analogous to the negative $U$ model
in which $\eta$ and CDW modes are conjugate
$ \left[ \eta^{\dagger} , \rho(Q) \right] = \Delta^{\dagger}$
and therefore coupled to each other.

{}From the fact that
the $\{ \eta^{\dagger},\eta , \rho(Q) \}$
operators for the negative $U$ Hubbard model
correspond to $\{ S_{+}, S_{-}, S_z(Q) \}$
for the Heisenberg model, it follows that oscillatons with
wavevectors close to $(\pi,\pi)$ will also be in one-to-one
correspondence for the two systems and will therefore have the same
dispersion.
So in the next section we briefly review what is known regarding
the excitations of the spin-wave modes of the
Heisenberg Hamiltonian in an external magnetic field,
Eq.~(\ref{heisenberg}).

\subsection{Dispersion for the Heisenberg model}
\label{sec:disp}

We consider a mean field spin--flop groundstate of the
Hamiltonian, Eq.~(\ref{heisenberg}), which has antiferromagnetic
order in the $x$--direction and is polarized in the $z$--direction
by the magnetic field $h$ so that
\begin{eqnarray}
< {\bf S}_i >~ = \frac{1}{2}
\left( \begin{array}{c} (-)^i \alpha \\ 0 \\ \beta
\end{array} \right).
\end{eqnarray}
As previously discussed, the antiferromagnetic order
corresponds to the superconducting order in the
negative $U$
system and the magnetic polarization to the deviation of the band
filling from half--filling. This means that temperature-wise
\mbox{$\alpha(T) \propto \Delta(T)$} and \mbox{$ \beta \propto 1-n =
  const$}.

If we now write the equations of motion for the individual spins
\begin{equation}
 \frac{ d {\bf S}_{i}} { d t } = i [ {\cal H}, {\bf S}_{i}]
\end{equation}
and linearize them around $< {\bf S}_i >$ we get
\begin{eqnarray}
\frac{ d~ \delta S_{i}^x}{ d t} &=&
\frac{J \beta}{2} \sum_{<j>} [ \delta S_j^y - \delta S_i^y ]
+ h \delta S_i^y \nonumber\\
\frac{ d~ \delta S_{i}^y}{ d t} &=&
(-)^i \frac{J \alpha }{2} \sum_{<j>}[ \delta S_j^z + \delta S_i^z ]
+\frac{J \beta}{2} \sum_{<j>} [ \delta S_i^x - \delta S_j^x ]
- h \delta S_i^x \nonumber\\
\frac{ d~ \delta S_{i}^z}{ d t} &=&
(-)^i \frac{J \alpha}{2} \sum_{<j>} [ - \delta S_j^y - \delta S_i^y ]
\label{spineqn}
\end{eqnarray}
where $<j>$ stands for the sites nearest to $i$.
We now take $M_{i}^z = (-)^i \delta S_{i}^z $ and $M_{i}^{x,y} =
\delta S_{i}^{x,y} $, and perform a Fourier transformation of
(\ref{spineqn}) to obtain
\begin{eqnarray}
- i \omega M_{k \omega}^x &=& - \frac{ J \beta}{2} ( 4 - \gamma_k ) M_{k
  \omega}^y + h M_{k \omega}^y \nonumber\\
- i \omega M_{k \omega}^y &=& \frac{ J \alpha}{2} ( 4 - \gamma_k ) M_{k
  \omega}^z + \frac{ J \beta}{2} ( 4 - \gamma_k ) M_{k
  \omega}^x - h M_{k \omega}^x \nonumber\\
- i \omega M_{k \omega}^z &=& - \frac{ J \alpha}{2} ( 4 + \gamma_k ) M_{k
  \omega}^y
\label{Meqn}
\end{eqnarray}
with $\gamma_k =\sum_{\bf a}e^{i{\bf k}{\bf a}}$.
Equation (\ref{Meqn}) may be solved giving the
dispersion relation\cite{RWhite}
\begin{equation}
\omega_k^2 = \frac{J^2 \alpha^2}{4} ( 16 - \gamma_k^2 ) + \left[ \frac{
  \beta J}{2} ( 4 - \gamma_k ) - h \right ]^2
\end{equation}
For small $k$'s the last expression may be expanded to give
\begin{equation}
\omega_k^2 = h^2 + ( 2 J^2 \alpha^2 - h \beta J ) \left[ ( k_x
  a )^2 + ( k_y a )^2 \right]
\label{smallk}
\end{equation}

Now we are in a position to discuss dispersion of the $\eta$-mode and
compare the Self-consistent Linear Response with the T-matrix
approximations.
Above $T_c$, there is no mixing between the particle hole and
the particle particle channels, a simple T-matrix calculation yields
the correct answer of a downward dispersion. This result agrees
with equation (\ref{smallk}) when we take $\alpha^2 \propto | \Delta|^2
= 0 $ above $T_c$. In the superconducting state a simple T-matrix
approximation
without taking into account the mixing is no longer valid. In fact, we
see from (\ref{smallk}) that the mixing term $ 2J^2 \alpha^2$ gives a
positive contribution to the dispersion. These two different terms
compete with each other, so that a positive dispersion can be obtained
at zero temperature for large $|U|$. On the other hand, for small
$|U|$, the second term $-h \beta J$ always wins over the first term $
2 J^2 \alpha^2$ and the dispersion is negative even at zero
temperature. This is consistent with the numerical results of the
Self-consistent Linear Response theory, presented in Section
\ref{sec:num}. From (\ref{smallk}), we also see that at large $|U|$,
as one rises the temperature from zero to $T_c$
the temperature dependence of $\alpha$ leads to a continious
transition from the positive dispersion to negative.

\section{SUMMARY AND CONCLUSIONS}
\label{sec:summary}

In this section we would like to summarize the results obtained
in the previous sections, and comment on their implications.
We see that the analytical approaches based on the $SO(4)$
symmetry property of the Hubbard model, the self-consistent linear
response theory and the numerical Monte Carlo simulations agree
with each other.  These calculations establish that the $\eta$
particle has the following properties:

1) It has charge quantum number two, and is a sharp excitation
of the system for a finite range of total momentum around $Q=(\pi,\pi)$,
and the mode disperses as the total momentum departs from $Q$.
The special role played by $Q$ is independent of the nesting property
of the fermi surface.

2) It has spin zero and energy $U-2\mu$.

3) Because it has charge two, it contributes to the particle--hole
fluctuation spectrum (in this case the density spectrum) only
when the superconducting pairing correlations are present.
The analytic calculations show that the intensity of the
peak onsets as the square of the superconducting order parameter.
The energy of the peak is unchanged as a function of temperature.

The value of present detailed study of the $\eta$ particle is to use it
to illustrate a mechanism with which particle particle excitations
can couple to the particle hole spectrum below $T_c$, and to test the
methodology of the theoretical approach.
We see that the
Self-consistent Linear Response method correctly takes into account
the contribution of the particle--particle excitation to the density
fluctuation spectrum, and agrees with both the exact theorems
derived from the $SO(4)$ Wards identities of the Hubbard model and
the numerical Monte Carlo simulation results.
This method is general,
and thus provides a useful starting point for investigating
the collective modes in the superconducting state.

In the next paper, we shall use the methodology developed and tested
in this paper to carry out an investigation of the $\pi$
particle of the positve $U$ Hubbard or the $t-J$ model, which has spin one
rather than zero, but shares many other properties of the $\eta$
particle. We shall discuss in detail the similarities and differences
of these two collective modes and compare our results with the neutron
scattering experiments.

\section{ACKNOWLEDGMENTS}
\label{sec:ack}

We would like to thank Prof. H. Fukuyama, R. Laughlin, J. R. Schrieffer
and C.N. Yang for useful discussions.
This work is supported by the NSF Materials Research Center at Stanford
University.
NB and DJS acknowledge support from the National Science Foundation
under grant DMR92--25027.
The computations were carried out at the San Diego Supercomputer Center.

\newpage
\section{Appendix A: Comparison with Equations of Motion Method}
\label{sec:append1}
In the equations of motion method we introduce
\begin{eqnarray}
\eta^{\dagger}_{kq}& = &c^{\dagger}_{k+q\uparrow}
c^{\dagger}_{-k\downarrow}\nonumber\\
\eta_{kq}& = & c_{-k-q\downarrow}
c_{k\uparrow}\nonumber\\
\rho_{kq}& = & \rho_{kq\uparrow}+\rho_{kq\downarrow}=c^{\dagger}_{k+q\uparrow}
c_{k\uparrow}+c^{\dagger}_{-k\downarrow} c_{-k-q\downarrow}\nonumber\\
\end{eqnarray}
and write the Heisenberg equations of motion for these
operators using (\ref{lrmodel}) as the Hamiltonian. These equations of
motion are then factorized in terms of the occupation numbers for the
electrons and BCS anomalous averages. The details of such computations
are presented extensively in the
literature\cite{Anderson1,Rickayzen,Schrieffer,vanderMarel}
and so we will only
outline the main
steps without going into details or discussing physical meaning of
the equations.

\begin{eqnarray}
\left[ {\cal H}, \rho_{pq\uparrow} \right] & = & \omega_{pq} \rho_{pq\uparrow}
-
( v_{p+q}^2 - v_p^2 ) U \left\{ \sum_k \rho_{kq\downarrow}
+\frac{1}{2} \phi_q
\right\} \nonumber\\
& + & u_p v_p U \sum_k \eta^{\dagger}_{kq} - u_{p+q} v_{p+q} U \sum_k \eta_{kq}
+ \Delta ( \eta^{\dagger}_{pq} - \eta_{pq} )
\\
\left[ {\cal H}, \rho_{pq\downarrow} \right] & =  &- \omega_{pq}
\rho_{pq\downarrow} +
( v_{p+q}^2 - v_p^2 ) U \left\{ \sum_k \rho_{kq\uparrow} +\frac{1}{2}
\phi_q \right\} \nonumber\\
 & + & u_{p+q} v_{p+q} U \sum_k \eta^{\dagger}_{kq} - u_{p} v_{p} U \sum_k
\eta_{kq}
+ \Delta ( \eta^{\dagger}_{pq} - \eta_{pq} )
\\
\left[ {\cal H}, \eta^{\dagger}_{pq} \right] & =  &\Omega_{pq}
\eta^{\dagger}_{pq} + U ( 1
- v_{p+q}^2 - v_p^2 ) \sum_k \eta^{\dagger}_{kq} \nonumber\\
 & + & U u_{p+q} v_{p+q} \left\{ \sum_k
\rho_{kq\uparrow} +\frac{1}{2} \phi_q \right\} + U u_p v_p \left\{
\sum_k \rho_{kq\downarrow} +\frac{1}{2} \phi_q \right\} + \Delta (
\rho_{pq\uparrow}+\rho_{pq\downarrow})
\\
\left[ {\cal H}, \eta_{pq} \right]  & = & - \Omega_{pq} \eta_{pq} - U ( 1
- v_{p+q}^2 - v_p^2 ) \sum_k \eta_{kq} \nonumber\\
 & - & U u_{p+q} v_{p+q} \left\{ \sum_k
\rho_{kq\uparrow} +\frac{1}{2} \phi_q \right\} - U u_p v_p \left\{
\sum_k \rho_{kq\downarrow} +\frac{1}{2} \phi_q \right\} - \Delta (
\rho_{pq\uparrow}+\rho_{pq\downarrow})
\end{eqnarray}

And we have already included the Hartree-Fock
terms into the renormalized chemical potential.
Following Anderson we take the second-order commutators

\begin{eqnarray}
\left[ {\cal H }, [ {\cal H } , \eta_{pq} - \eta^{\dagger}_{pq} ]
\right] & =  &
( \Omega_{pq}^2 + 4 \Delta^2 ) ( \eta_{pq} - \eta^{\dagger}_{pq} ) - 2
\Delta \omega_{pq} ( \rho_{pq\uparrow}-\rho_{pq\downarrow} )
\nonumber\\
 & +  & U \Omega_{pq} ( 1 - v_{p+q}^2 - v_p^2 ) \sum_k ( \eta_{kq} -
\eta^{\dagger}_{kq} )
\nonumber\\
& + & 2 U \Delta ( u_p v_p + u_{p+q} v_{p+q} ) \sum_k ( \eta_{kq} -
\eta^{\dagger}_{kq} )
\nonumber\\
& - & 2 U \Delta ( v_{p+q}^2 - v_p^2 ) \sum_k (\rho_{kq\uparrow} -
\rho_{kq\downarrow} )
\nonumber\\
 & -  & U ( u_p v_p + u_{p+q} v_{p+q} ) \left [
{\cal H}, \sum_k ( \rho_{kq\uparrow}+\rho_{kq\downarrow} ) \right ]
\nonumber\\
& - & U ( 1 - v_{p+q}^2 - v_p^2 ) \left [
{\cal H}, \sum_k ( \eta_{kq}+\eta^{\dagger}_{kq} ) \right ]
\label{b-b+}
\\
\left[ {\cal H }, [ {\cal H } , \eta_{pq} + \eta^{\dagger}_{pq} ]
\right]  & = &
 \Omega_{pq}^2 ( \eta_{pq} + \eta^{\dagger}_{pq} ) + 2
\Delta \Omega_{pq} ( \rho_{pq\uparrow}+\rho_{pq\downarrow} )
\nonumber\\
 & +  & U \Omega_{pq} ( 1 - v_{p+q}^2 - v_p^2 ) \sum_k ( \eta_{kq} +
\eta^{\dagger}_{kq} )
\nonumber\\
& + & U \Omega_{pq} ( u_p v_p + u_{p+q} v_{p+q} )  \left\{ \sum_k (
\rho_{kq\uparrow}+\rho_{kq\downarrow} ) + \phi_q \right\}
\nonumber\\
 & -  & U ( 1 - v_{p+q}^2 - v_p^2 ) \left [
{\cal H}, \sum_k ( \eta_{kq}-\eta^{\dagger}_{kq} ) \right ]
\label{b+b+}
\\
\left[ {\cal H }, [ {\cal H } , \rho_{pq\uparrow}-\rho_{pq\downarrow}]
\right ] & = &
  \omega_{pq}^2 (\rho_{pq\uparrow}-\rho_{pq\downarrow}) -
2 \Delta \omega_{pq} ( \eta_{pq} - \eta^{\dagger}_{pq} )
\nonumber\\
& - &U \omega_{pq} ( u_p v_p + u_{p+q} v_{p+q} ) \sum_k ( \eta_{kq} -
\eta^{\dagger}_{kq} )
\nonumber\\
& + & U \omega_{pq} ( v_{p+q}^2 - v_p^2 ) \sum_k (
\rho_{kq\uparrow} - \rho_{kq\downarrow} )
\nonumber\\
& - & U ( v_{p+q}^2 - v_p^2 ) \left [
{\cal H}, \sum_k ( \rho_{kq\uparrow}+\rho_{kq\downarrow} ) \right ]
\nonumber\\
& + & U ( u_p v_p - u_{p+q} v_{p+q} ) \left [
{\cal H}, \sum_k ( \eta_{kq}+\eta^{\dagger}_{kq} ) \right ]
\label{up-down}
\\
\left[ {\cal H }, [ {\cal H } , \rho_{pq\uparrow}+\rho_{pq\downarrow}]
\right ] & = &
  ( \omega_{pq}^2 + 4 \Delta^2 ) (\rho_{pq\uparrow}+\rho_{pq\downarrow}) +
2 \Delta \Omega_{pq} ( \eta_{pq} + \eta^{\dagger}_{pq} )
\nonumber\\
& + & U \omega_{pq} ( u_p v_p - u_{p+q} v_{p+q} ) \sum_k ( \eta_{kq} +
\eta^{\dagger}_{kq} )
\nonumber\\
& + & 2 \Delta U (1 - v_{p+q}^2 - v_p^2  ) \sum_k ( \eta_{kq} +
\eta^{\dagger}_{kq} )
\nonumber\\
& + & 2 \Delta U ( u_p v_p + u_{p+q} v_{p+q} ) \left\{ \sum_k (
\rho_{kq\uparrow}+\rho_{kq\downarrow} ) + \phi_q \right\}
\nonumber\\
& + & U ( v_{p+q}^2 - v_p^2 ) \left [
{\cal H}, \sum_k ( \rho_{kq\uparrow}-\rho_{kq\downarrow} ) \right ]
\nonumber\\
& - & U ( u_p v_p + u_{p+q} v_{p+q} ) \left [
{\cal H}, \sum_k ( \eta_{kq}-\eta^{\dagger}_{kq} ) \right ]
\label{up+down}
\end{eqnarray}

Now we take $ \left[ {\cal H } , f \right ] = - \omega
f $, where $f$ can be any of $ \rho, \eta^{\dagger}$ or $ b $, combine
equations (\ref{b-b+}) with (\ref{up-down})  and
(\ref{b+b+}) with (\ref{up+down}) such that unphysical poles
$(\omega^2-\theta_{pq}^2)$ cancel out and get after a few
straightforward transformations:

\begin{eqnarray}
(\omega^2-\nu_{pq}^2) \rho_{pq} & = &
U \Omega_{pq} (u_p v_p +u_{p+q} v_{p+q}) \sum_{k}
(\eta^{\dagger}_{kq}+\eta_{kq}) \nonumber\\
 & -  & U \omega (u_p v_p +u_{p+q} v_{p+q}) \sum_{k}
(\eta^{\dagger}_{kq}-\eta_{kq}) \nonumber\\
 & + & U  \nu_{pq} ( v_p u_{p+q}+v_{p+q}u_p)^2 ( \sum_{k}\rho_{kq}+\phi_q)
\label{ro}
\\
(\omega^2-\nu_{pq}^2)~ (\eta^{\dagger}_{pq}+\eta_{pq})& = &
U \Omega_{pq} (1-v_{p+q}^2-v_{p}^2)
\sum_{k} (\eta^{\dagger}_{kq}+\eta_{kq}) \nonumber\\
& - & U \omega (1-v_{p+q}^2-v_{p}^2)
\sum_{k}( \eta^{\dagger}_{kq}-\eta_{kq})  \nonumber\\
& + & U \Omega_{pq} (u_p v_p +u_{p+q} v_{p+q})
( \sum_{k}\rho_{kq}+\phi_q )
\label{z}
\\
(\omega^2-\nu_{pq}^2) ~(\eta^{\dagger}_{pq}-\eta_{pq})  = &
- & U \omega (1-v_{p+q}^2-v_{p}^2) \sum_{k} (
\eta^{\dagger}_{kq}+\eta_{kq} )
\nonumber\\
& + & U \nu_{pq} (u_p u_{p+q} +v_p v_{p+q})^2
\sum_{k}( \eta^{\dagger}_{kq}-\eta_{kq}) \nonumber\\
& - & U \omega (u_p v_p+u_{p+q} v_{p+q})
( \sum_{k} \rho_{kq}+\phi_q )
\label{y}
\end{eqnarray}
One can see that after we divide
equations (\ref{ro}) - (\ref{y}) by $(\omega^2-\nu_{pq}^2)$ and sum
over $p$ we recover the formulas of the Linear Response. The equations
of motion method is very often more convenient because one gets
equations before they were summed over momentum and this gives some
additional freedom in manipulating them. So, for example it is much
easier to see the $\eta$-pair energy-eigenvalue or to compute the
residue of $D(q,\omega)$ at $\omega_0$ in equations(\ref{ro}) -
(\ref{y}) than in (\ref{matrix}) - (\ref{chi}). The same time the Linear
Response Theory has an advantage of being more intuitively clear and
transparent.

\newpage
\section{Appendix B: Diagrammatic equations}

In this appendix we show the diagrams that correspond to equations
(\ref{b+freq}) - (\ref{rofreq}).

The first equation describes scattering in the particle-particle
channel and should be compared to (\ref{b+freq}).
\vspace{1 cm}

\begin{figure}[hbt]
    \ifundefined{epsffile}\relax\else
\centerline{ \epsffile[ 154 534 444 732]{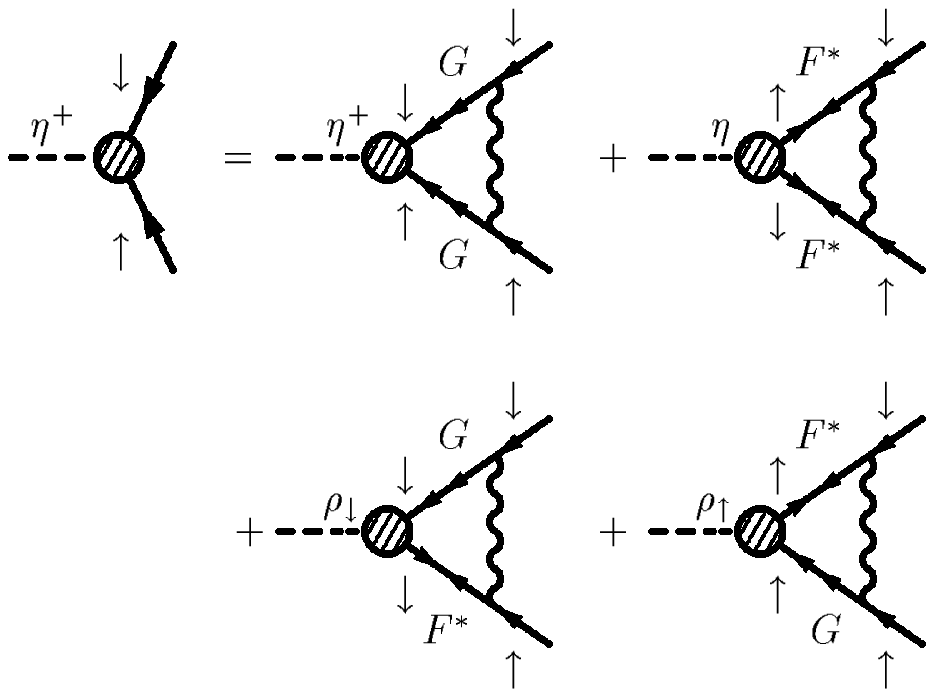}}\fi
\end{figure}

This next equation is for scattering in the hole-hole channel
and corresponds to (\ref{b-freq}).
\vspace{1 cm}

\begin{figure}[hbt]
    \ifundefined{epsffile}\relax\else
    \centerline{ \epsffile[ 154 534 444 732 ]{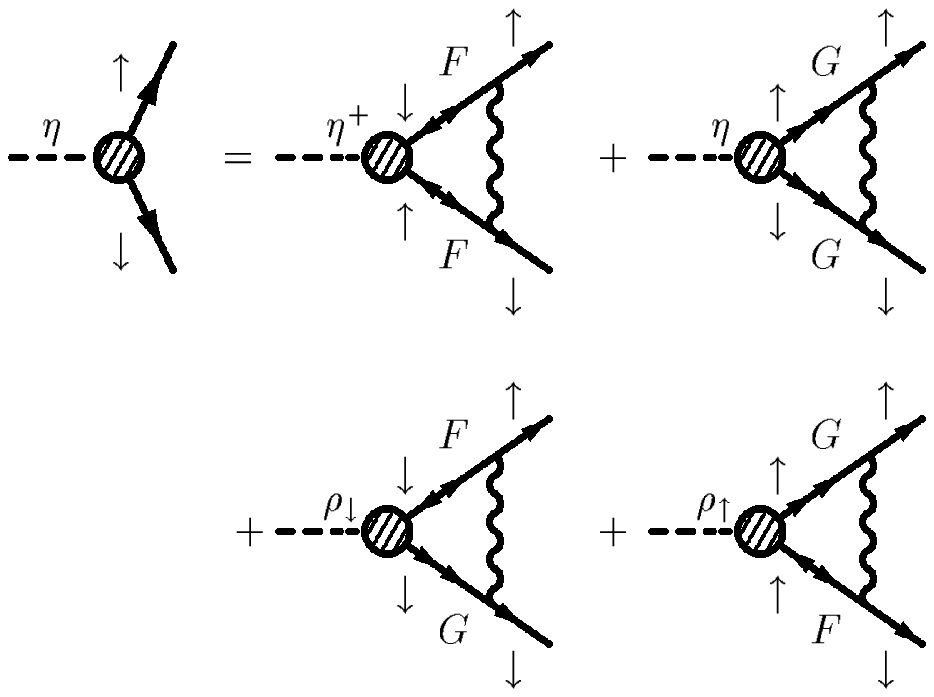}}\fi
\end{figure}

\newpage
The two following equations are for scattering in the particle-hole
channels with spins up or down. Their sum gives (\ref{rofreq}).

\vspace{2 cm}
Spin up:

\begin{figure}[hbt]
    \ifundefined{epsffile}\relax\else
\centerline{ \epsffile[20 547 580 721]{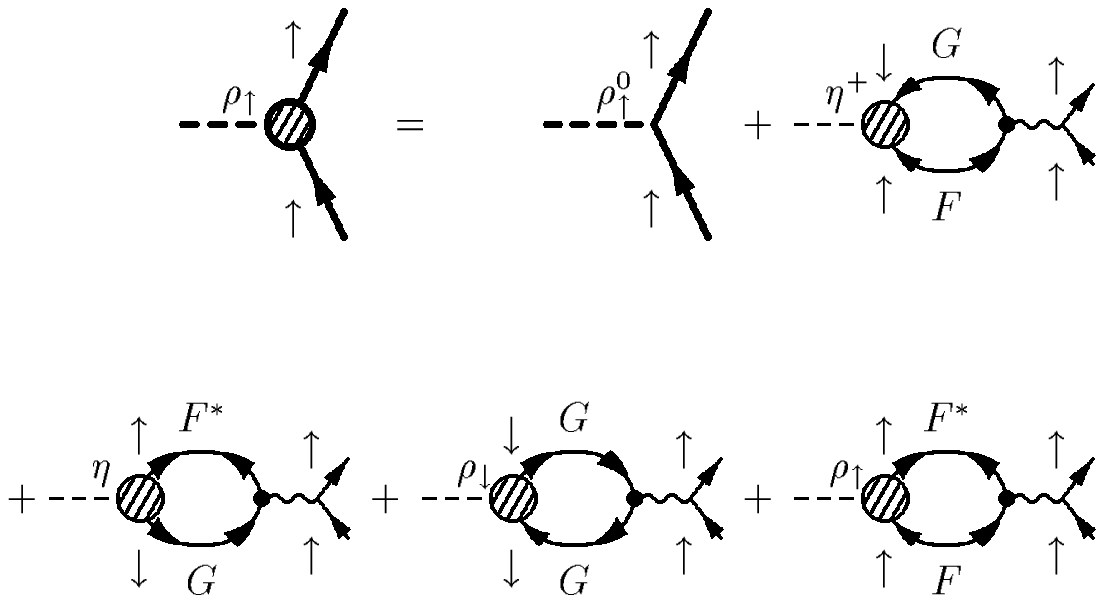}}\fi
\end{figure}

\vspace{2 cm}
Spin down:

\begin{figure}[hbt]
    \ifundefined{epsffile}\relax\else
\centerline{  \epsffile[20 547 550 721]{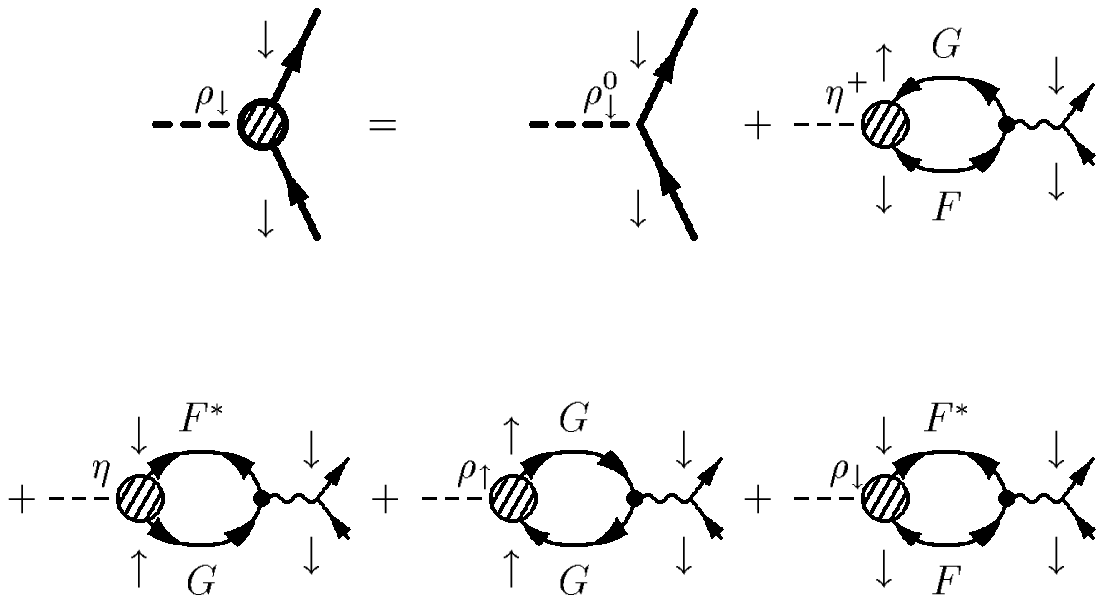}}\fi
\end{figure}


\begin{thebibliography}{10}

\bibitem{Yang-Zhang}
C. Yang and S. Zhang, Modern Physics Letters B {\bf 4},  759  (1990).

\bibitem{Zhang1}
S. Zhang, Phys. Rev. Lett. {\bf 65},  120  (1990).

\bibitem{Zhang2}
S. Zhang, International Journal of Modern Physics B {\bf 5},  153  (1991).

\bibitem{Anderson1}
P. Anderson, Phys. Rev. {\bf 112},  1900  (1958).

\bibitem{Rickayzen}
G. Rickayzen, Phys. Rev. {\bf 115},  795  (1959).

\bibitem{Schrieffer}
J. Schrieffer, {\em Theory of Superconductivity} (Benjamin/Cummings Publishing
  Co., Reading, Massachusetts, 1971).

\bibitem{Bardasis}
A. Bardasis and J. Schrieffer, Phys. Rev. {\bf 116},  235  (1987).

\bibitem{vanderMarel}
van~der Marel, Phys. Rev. B {\bf 51},  1147  (1995).

\bibitem{Kostyrko2}
T. Kostyrko and R. Micnas, Phys. Rev. B {\bf 46},  11 025  (1992).

\bibitem{Grenoble1}
J. Rossat-Mignod {\it et~al.}, Physica C {\bf 185-189},  86  (1991).

\bibitem{Grenoble}
J. Rossat-Mignod {\it et~al.}, Physica C {\bf 235-240},  1687  (1994).

\bibitem{Mook1}
H. Mook {\it et~al.}, Phys. Rev. Lett. {\bf 70},  3490  (1994).

\bibitem{Keimer}
H.~F. Fong {\it et~al.}, Phys. Rev. Lett. {\bf 75},  316  (1995).

\bibitem{pi}
E. Demler and S. Zhang, Phys. Rev. Lett. {\bf 75},  4126  (1995).

\bibitem{Yang}
C. Yang, Phys. Rev. Lett. {\bf 63},  2144  (1989).

\bibitem{Machida}
K. Machida, K. Nokura, and T. Matsubara, Phys. Rev. Lett. {\bf 44},  821
  (1980).

\bibitem{QMC}
S. White {\it et~al.}, Phys. Rev. B {\bf 40},  506  (1989).

\bibitem{maxent1}
R. Silver, D. Sivia, and J. Gubernatis, Phys. Rev. B {\bf 41},  2380  (1990).

\bibitem{maxent2}
S. White, Phys. Rev. B {\bf 41},  2380  (1990).

\bibitem{negU1}
R. Scalettar {\it et~al.}, Phys. Rev. Lett. {\bf 62},  1407  (1989).

\bibitem{negU2}
A. Moreo and D. Scalapino, Phys. Rev. Lett. {\bf 66},  946  (1991).

\bibitem{SWZ}
D. Scalapino, S. White, and S. Zhang, Phys. Rev. B {\bf 47},  7995  (1993).

\bibitem{Mele}
E. Mele, Solid State Communications {\bf 79},  515  (1991).

\bibitem{Shiba}
H. Shiba, Prog. Theor. Phys {\bf 48},  2171  (1971).

\bibitem{Hirsch}
J.~E. Hirsch, Phys. Rev. B {\bf 31},  4403  (1985).

\bibitem{RWhite}
R.~M. White, {\em Quantum Theory of Magnetism} (McGraw--Hill, New York, 1970).

\end{thebibliography}
\end{document}